\documentclass[]{elsarticle}

\usepackage{lineno,hyperref}
\modulolinenumbers[5]

\usepackage[margin=2cm]{geometry}
\usepackage[utf8]{inputenc}
\usepackage{url}
\usepackage{array}
\usepackage{amsmath,amssymb,amsfonts}
\usepackage{mathtools}
\usepackage{algorithm}
\usepackage{algpseudocode}
\usepackage{tikz}
\usepackage{graphicx}
\usepackage{subcaption}
\usepackage[export]{adjustbox}
\usepackage{bm}
\usepackage{xcolor}
\usepackage{tabularx}

\journal{ }

\usepackage{numcompress}

\begin{document}

\begin{frontmatter}

\title{Digital Blood in Massively Parallel CPU/GPU Systems for the Study of Platelet Transport}

\author[UniGe]{Christos Kotsalos \corref{correspondingauthor}}
\ead{christos.kotsalos@unige.ch}

\author[UniGe]{Jonas Latt}

\author[UniGe]{Joel Beny}

\author[UniGe]{Bastien Chopard}

\address[UniGe]{Computer Science Department, University of Geneva, 7 route de Drize, CH-1227 Carouge, Switzerland}

\cortext[correspondingauthor]{Corresponding Author}

\begin{abstract}
We propose a highly versatile computational framework for the simulation of cellular blood flow focusing on extreme performance without compromising accuracy or complexity. The tool couples the lattice Boltzmann solver Palabos for the simulation of the blood plasma, a novel finite element method (FEM) solver for the resolution of the deformable blood cells, and an immersed boundary method for the coupling of the two phases. The design of the tool supports hybrid CPU-GPU executions (fluid, fluid-solid interaction on CPUs, the FEM solver on GPUs), and is non-intrusive, as each of the three components can be replaced in a modular way. The FEM-based kernel for solid dynamics outperforms other FEM solvers and its performance is comparable to the state-of-the-art mass-spring systems. We perform an exhaustive performance analysis on Piz Daint at the Swiss National Supercomputing Centre and provide case studies focused on platelet transport. The tests show that this versatile framework combines unprecedented accuracy with massive performance, rendering it suitable for the upcoming exascale architectures.
\end{abstract}


\end{frontmatter}

\let\today\relax 

\section{Background}
Blood flow plays an important role in most of the fundamental functions of living organisms. Blood transports oxygen, nutrients, waste products, infectious parasites, tumour cells, to name a few, to tissues and organs. Despite remarkable advances in experimental techniques \cite{Tomaiuolo2014BiomechanicalMicrofluidics}, the type and detail of information provided remains limited. In the last two decades, computational tools for the direct numerical simulation of cellular blood flow have been developed. They complement experiments and have become an essential tool for in-depth investigations. These tools \cite{Freund_2014} have been used to study various poorly understood phenomena such as the non-Newtonian viscosity of blood \cite{Zavodszky2017CellularCells}, the thrombus formation \cite{Fogelson_Review_2015}, the F\r{a}hr{\ae}us effect, the characteristics of the cell free layer \cite{Vahidkhah_2014}, and the red blood cell (RBC) enhanced shear-induced diffusion of platelets \cite{Mehrabadi_2015}, \cite{Zhao_PLTs_2011}. Apart from physiological conditions, numerical tools have significantly assisted the understanding of pathological conditions \cite{Fedosov_2011a}, \cite{Li_2017}, \cite{Chang_Karniadakis_2018}, as they offer a controlled environment for testing a large number of parameters and classifying their effect on blood rheology. With the occurrence of more mature tools, there is an increased focus on developing/ analysing lab-on-chip systems \cite{Rossinelli_2015}, \cite{Kruger_2014} and drug delivery systems \cite{Gupta_2016}, \cite{Vahidkhah_2015}. Despite such advances, we believe that there is a tremendous space for improvement in terms of fidelity, high-performance and clinically relevant scales.

Blood is a complex suspension of RBCs, white blood cells and platelets, submerged in a Newtonian fluid, the plasma. At $35-45\%$ hematocrit, RBCs occupy a substantial volume fraction of the blood and have therefore an important impact on the blood rheology. The accurate modelling of the collective transport of cells in the plasma is of paramount importance, since it can help decipher various \emph{in vivo} and \emph{in vitro} phenomena. A single $mm^3$ of blood (almost a blood drop) contains a few million RBCs, a few hundred thousand platelets and a few thousand white blood cells. Thus, it is extremely challenging to develop a tool capable of simulating blood at cellular level for clinically relevant applications, using high fidelity models and utilising a reasonably limited amount of computational resources and time.

The absence of such a universal cellular blood flow computational tool constitutes the motivation behind the suggested framework. A universal framework should fulfil a number of criteria, such as generality, robustness, accuracy, performance and modularity. The criteria of generality, robustness and accuracy are addressed in our first description of the tool proposed in Kotsalos \emph{et al.} \cite{Kotsalos_JCP_2019_npFEM}. In this work we complete the framework by introducing an integrated environment that obeys all the criteria towards a universal computational tool for digital blood. Our framework is tailored for the fastest supercomputers (namely hybrid CPU/GPU clusters, which are well represented at the high end of the TOP500 list) and it is ready to be hosted in the upcoming exascale machines. Moreover, the suggested tool, even if it uses state-of-the-art numerical techniques, is not monolithically built upon them: the structural solver for the blood constituents can easily be replaced by another one, such as one based on discrete element models. Similarly, the lattice Boltzmann flow solver could be replaced by another option, which however needs to be similarly parallelisable through domain decomposition and allow interaction with solid particles through an immersed boundary method. In the last decade there are many groups working on HPC-capable cellular blood flow tools \cite{Rahimian_2010}, \cite{Peters_2010}, \cite{Bernaschi_2011}, \cite{Xu_2013_RBCs}, \cite{Xu_2017_RBCs}, \cite{Rossinelli_2015} dealing with problems of increased complexity, which do however not reach the goal of a computational framework that fulfils simultaneously all above-mentioned criteria.

Our team (Palabos development group \cite{PalabosArticle}, \cite{PalabosWebsite}) has developed a numerical method and a high-performance computing (HPC) software tool for ab initio modelling of blood. The framework models blood cells like red blood cells and platelets individually, including their detailed non-linear elastic properties and a complex interaction between them. The project is particularly challenging because the large number of blood constituents (up to billions) stands in contrast with the high computational requirement of individual elements. While classical approaches address this challenge through simplified structural modelling of deformable RBCs (e.g. through mass-spring systems) \cite{Dupin2007ModelingDimensions}, \cite{Fedosov2010ARheologydynamics}, \cite{Reasor2011CouplingFlow}, \cite{Rossinelli_2015}, \cite{Zavodszky2017CellularCells}, the present framework guarantees accurate physics and desirable numerical properties through a fully-featured FEM model \cite{Kotsalos_JCP_2019_npFEM}. The required numerical performance is achieved through a hybrid implementation, using CPUs (central processing units) for the blood plasma and GPUs (graphics processing units) for the blood cells. The developed numerical framework is intended to grow to be a general-purpose tool for first-principle investigation of blood properties and to provide an integrative and scalable HPC framework for the simulation of blood flows across scales.

The present work is organised as follows: In section \ref{sec:Methods}, we  present the structure of our computational tool and the basic underlying methods. In section \ref{sec:ResAndDisc}, we provide a performance analysis on the Piz Daint supercomputer and various case studies of platelet transport.

\section{Methods} \label{sec:Methods}
\subsection{Computational Framework}
The understanding and deciphering of a complex transport phenomenon like the movement of platelets requires the deployment of high fidelity direct numerical simulation tools that resolve the cellular nature of blood. Platelets are submerged in the blood plasma and collide continuously with the RBCs, also known as erythrocytes, that are present in much larger quantities. To capture accurately their trajectories and understand the driving mechanisms behind their motion, we propose a modular and generic high-performance computational framework capable of resolving fully 3D blood flow simulations. The computational tool is built upon three modules, namely the fluid solver, the solid solver and the fluid-solid interaction (FSI).

The fluid solver is based on the lattice Boltzmann method (LBM) and solves indirectly the weakly compressible Navier-Stokes equations. The 3D computational domain is discretised into a regular grid with spacing $\Delta x$ in all directions. For this study, we use the D3Q19 stencil, with the BGK collision operator and non-dimensional relaxation time $\tau = 2$ (higher $\tau$ gives higher $\Delta t$). The time step is determined through the formula $\nu = C_s^2 \left ( \tau - 1/2 \right ) \Delta t$, where the fluid speed of sound is $C_s = \Delta x/ \Delta t \sqrt{3}$ and $\nu$ the kinematic viscosity. Furthermore, external forcing terms (like the FSI force $f_{imm}$) can be incorporated in the LBM through a modification of the collision operator using the Shan-Chen forcing scheme \cite{Shan1993LatticeComponents}. More information on LBM can be found in \cite{Feng_LBM_ComputationalIssues}, \cite{Kruger2017TheMethod}, \cite{KrugerThesis}.

The solid solver is based on the recently introduced nodal projective finite elements method (npFEM) by Kotsalos \emph{et al.} \cite{Kotsalos_JCP_2019_npFEM}, which offers an alternative way of describing elasticity. The npFEM framework is a mass-lumped linear FE solver that resolves both the trajectories and deformations of the blood cells with high accuracy. The solver has the capability of capturing the rich and non-linear viscoelastic behaviour of red blood cells as shown and validated in \cite{Kotsalos_JCP_2019_npFEM}. The platelets are simulated as nearly-rigid bodies by modifying the stiffness of the material. The implicit nature of the npFEM solver renders it capable of resolving extreme deformations with unconditional stability for arbitrary time steps. Even though the solver is based on FEM and an implicit integration scheme, its performance is very close to the widely used mass-spring systems \cite{Fedosov2010ARheologydynamics}, \cite{Zavodszky2017CellularCells}, outperforming them in robustness and accuracy \cite{Kotsalos_JCP_2019_npFEM}. Regarding the blood cell viscoelastic behaviour, the solver uses a two-way approach to handle the response of the cell to the imposed loads over time (Rayleigh and position-based dynamics damping \cite{Kotsalos_JCP_2019_npFEM}). It should be pointed out that the interior fluid of the cell is implicitly taken into account, as its incompressibility contributes to the potential energy of the body and its viscosity augments the viscosity of the membrane. Furthermore, the force on the bodies is derived from the hydrodynamic stress tensor by considering the lattice points at the exterior of the bodies. A complete presentation of npFEM can be found in Kotsalos \emph{et al.} \cite{Kotsalos_JCP_2019_npFEM}.


The fluid-solid interaction is realised by the immersed boundary method (IBM) and more specifically by the multi-direct forcing scheme proposed by Ota \emph{et al.} \cite{Ota2012LiftSimulations} (with minor modifications, see supplementary material). The IBM imposes a no-slip boundary condition, so that each point of the surface and the ambient fluid moves with the same velocity. The advantage of the IBM is that the fluid solver does not have to be modified except for the addition of a forcing term $f_{imm}$. Moreover, the deformable body and its discrete representation do not need to conform to the discrete fluid mesh, which leads to a very efficient fluid-solid coupling. The exchange of information from the fluid mesh to the Lagrangian points of the discrete bodies and vice versa is realised through interpolation kernels with finite support. The $\phi_4$ kernel \cite{Mountrakis_2017_IBM} is used throughout the simulations of the current study. The IBM is an iterative algorithm where the force estimate on a Lagrangian point is computed by the difference of the vertex velocity and the velocity interpolated by the surrounding fluid nodes. Then, this force is spread onto the fluid nodes ($f_{imm}$) surrounding the Lagrangian point and the correction affects the collision operator of the LBM. This interpolation between the membranes and the fluid is repeated for a finite amount of steps. For the simulations shown in this article, just one iteration suffices for the required accuracy.

A brief but more instructive overview of the methods presented above can be found in the supplementary material.

\subsection{Towards stable and robust FSI}
There exist two main ways to realise fluid-solid interaction, which are monolithic and modular respectively. The former describes the fluid and solid phases through one system of equations and both are solved with a single solver, using as well the same discretisation. Examples include tools that use dissipative particle dynamics to resolve both the fluid and solid. An advantage of the monolithic approach is the consistency of the coupling scheme, which leads to more numerically stable solutions. The main disadvantage is that a single solver potentially falls short of satisfactorily addressing all the physics present in a complex phenomenon. In the modular approach, there is the freedom to choose well optimised solvers to address the different phases. However, the consistent coupling of the solvers becomes a major challenge, especially when the discretisation (space \& time) is non-conforming. Our computational framework uses different spatial and time resolutions for the fluid and solid phases. For example, the solid solver is executed every two iterations (at least), which could potentially introduce coupling instabilities. The instabilities arise mainly from under-resolution and from integration schemes that do not conserve energy and momenta (linear/ angular) exactly, thus leading to spirally increasing energy oscillations between the solvers. The remedies suggested below are tailored to the specific framework, but could potentially give useful hints for other implementations.

The IBM requires a match between the lattice spacing $\Delta x$ and the average edge length $\bar{l}$ of the discretised membranes (triangulated surfaces). The value of the mesh ratio $\bar{l}/\Delta x$ appears to play a minor role as long as it is comprised in the range  $[0.5, 1.8]$ \cite{KrugerThesis}. A RBC discretised with 258 surface vertices exhibits a ratio $\bar{l}/\Delta x \sim 1.6$ with a lattice spacing of $0.5 ~\mu m$. For low shear rates, this requirement can be further relaxed.

An accurate evaluation of the external forces acting on the immersed boundaries plays a critical role to achieve a consistent coupling. For higher accuracy we use the hydrodynamic stress tensor $\bm{\sigma}$ projected onto the surface normals instead of the native force term produced by the IBM. Furthermore, compatible with the aim to disregard the interior fluid of the blood cells, we found out that the most stable force evaluation scheme comes from measuring $\bm{\sigma}$ at the exterior most distant site from the Lagrangian point contained within the interpolation kernel.

A meticulous handling of the near-contact regions is deemed highly critical to suppress instabilities. The first step of our procedure consists of searching for Lagrangian points belonging to bodies other than the investigated one that are contained within the interpolation kernel of the current point. If there are no ``foreign'' particles in the kernel then no modification is needed. It is then assumed that the interaction physics is appropriately resolved by the fluid field in between bodies. Otherwise, the collision framework takes over, since the evaluation of $\bm{F}_{ext}$ is ``contaminated'' by the interior fluid of a neighbouring body. Subsequently, the forces on the Lagrangian point from the fluid are disregarded and a collision force, coming from a quadratic potential energy \cite{Kotsalos_JCP_2019_npFEM}, is used instead. This technique is named by us particle in kernel (PIK) and resolves very accurately colliding bodies (more in supplementary material). We would like to highlight that the actual IBM algorithm is not affected by the PIK technique.

The selected IBM version \cite{Ota2012LiftSimulations} starts from an estimate of a force at the Lagrangian points required to enforce a no-slip condition. This force is spread into the neighbourhood of the points to communicate the constraints of the solid boundary to the fluid. The force estimate is proportional to the difference of the vertex velocity (as computed by the npFEM solver) and the velocity interpolated by the surrounding fluid nodes. The component that can be controlled in the above procedure is the npFEM vertex velocity which, if it exceeds a value of $U^{\ast}_{max}$, is truncated towards this threshold. The constant $U^{\ast}_{max}$ can be comprised between $[0.03,0.1]$ \cite{Kruger2017TheMethod} and is related to the fact that the simulated Mach number $Ma$ should be $\ll 1$, since LBM errors increase dramatically at high $Ma$. This velocity capping proves to be very stabilising when necessary. If the classic IBM \cite{Peskin1972FlowMethod} is used, then a force capping has the aforementioned stabilising effect.

Once the force $\bm{F}_{ext}$ is computed, a median filtering in the one-ring neighbourhood of every Lagrangian point attenuates force spikes that could result in energy oscillations.

\subsection{High Performance Computing (HPC) Design}
Direct numerical simulations of cellular blood flow are pushing the computational limits of any modern supercomputer, given the complexity of the underlying phenomena. The amount of unknowns per second varies from millions to trillions \cite{Rossinelli_2015}, and the proposed computational framework is built with genericity, modularity and performance in mind, able to tackle problems in the whole range of unknowns. This computational tool is orchestrated by Palabos \cite{PalabosArticle}, \cite{PalabosWebsite}, which is responsible for data decomposition and for the communication layer. Palabos (for {\bf Pa}rallel {\bf La}ttice {\bf Bo}ltzmann {\bf S}olver) is an open-source library for general-purpose computational fluid dynamics, with a ``kernel'' based on the lattice Boltzmann method. Palabos is written in C\texttt{++} with extensive use of the Message Passing Interface (MPI) and with proven HPC capability, particularly in the domain of computational biomedicine \cite{Mountrakis2015ParallelFramework}, \cite{Zavodszky2017Hemocell:Library}, \cite{Tan_2018_PALABOS_LAMMPS}. On top of Palabos core library, we have developed the npFEM solver, which is written in C\texttt{++} and CUDA, a general purpose parallel computing platform and programming model for NVIDIA GPUs and it is derived from the open-source library ShapeOp \cite{ShapeOpWebSite}. There are two active branches of the npFEM library, a CPU-only implementation and a full GPU implementation leveraging NVIDIA GPUs. The GPU parallelization strategy is based on the idea of using one CUDA-thread per Lagrangian point and one CUDA-block per blood cell. This is feasible since the most refined blood cell model has less points (discretised membrane) than the maximum allowed number of threads per block (hardware constraint). Keeping all points of a cell within a single CUDA-block allows us to compute the entire solver time step in one CUDA-kernel call, thus making good use of cache and shared memory \cite{Beny_2019}.

\begin{figure}[h]
    \centering
    \includegraphics[scale=0.45]{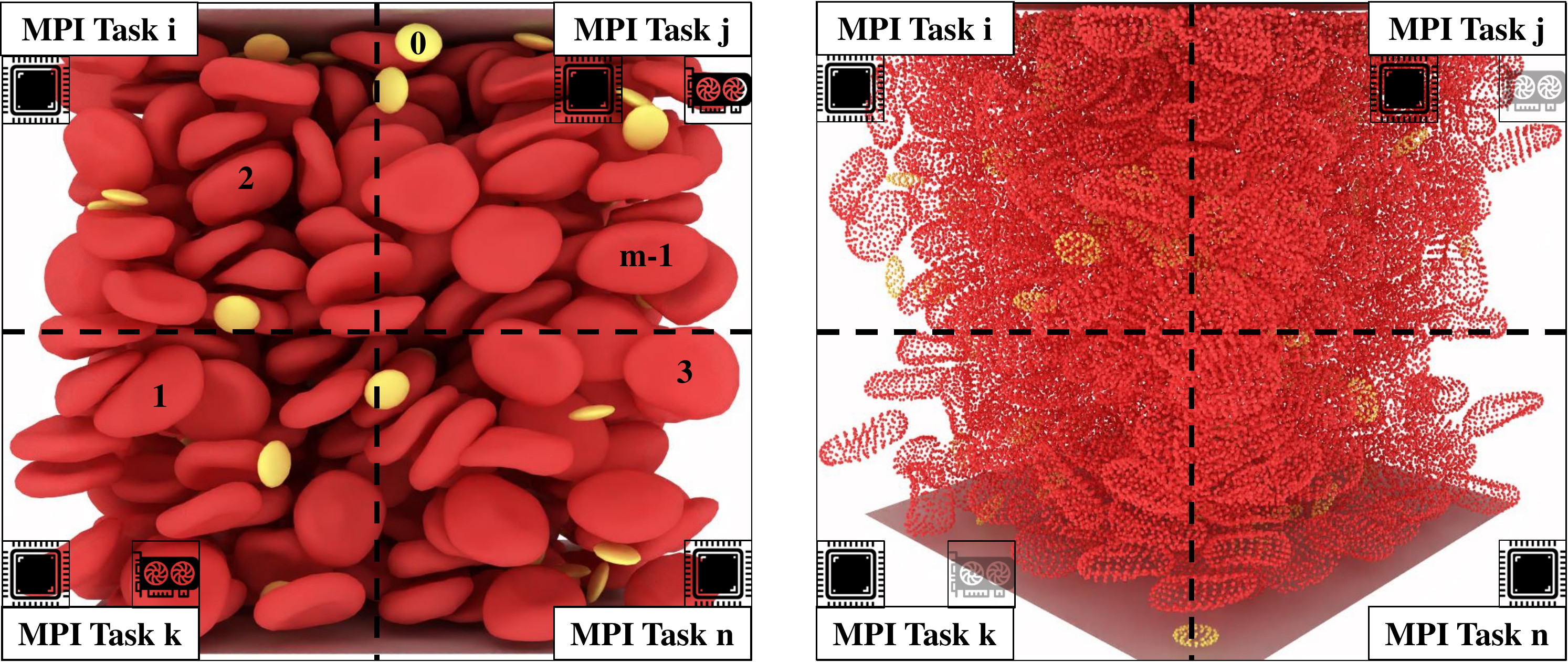}
    \caption{Load balancing of the fluid and solid phases for a modern CPU/GPU computing system. Cubic domain at $35\%$ hematocrit with RBCs and platelets. The immersed bodies exist both as plain Lagrangian points (positions and velocities for the IBM) and augmented Lagrangian points (positions, velocities and other properties relevant only to the solid solver). The LBM and IBM are executed on CPUs, while the npFEM on GPUs (hybrid version of the framework). In most cases, the number of GPUs is way less than the available MPI tasks, see Piz Daint with a ratio 1:12.}
    \label{fig:LoadBalancing}
\end{figure}

\begin{figure}[H]
    \centering
    \includegraphics[scale=0.55]{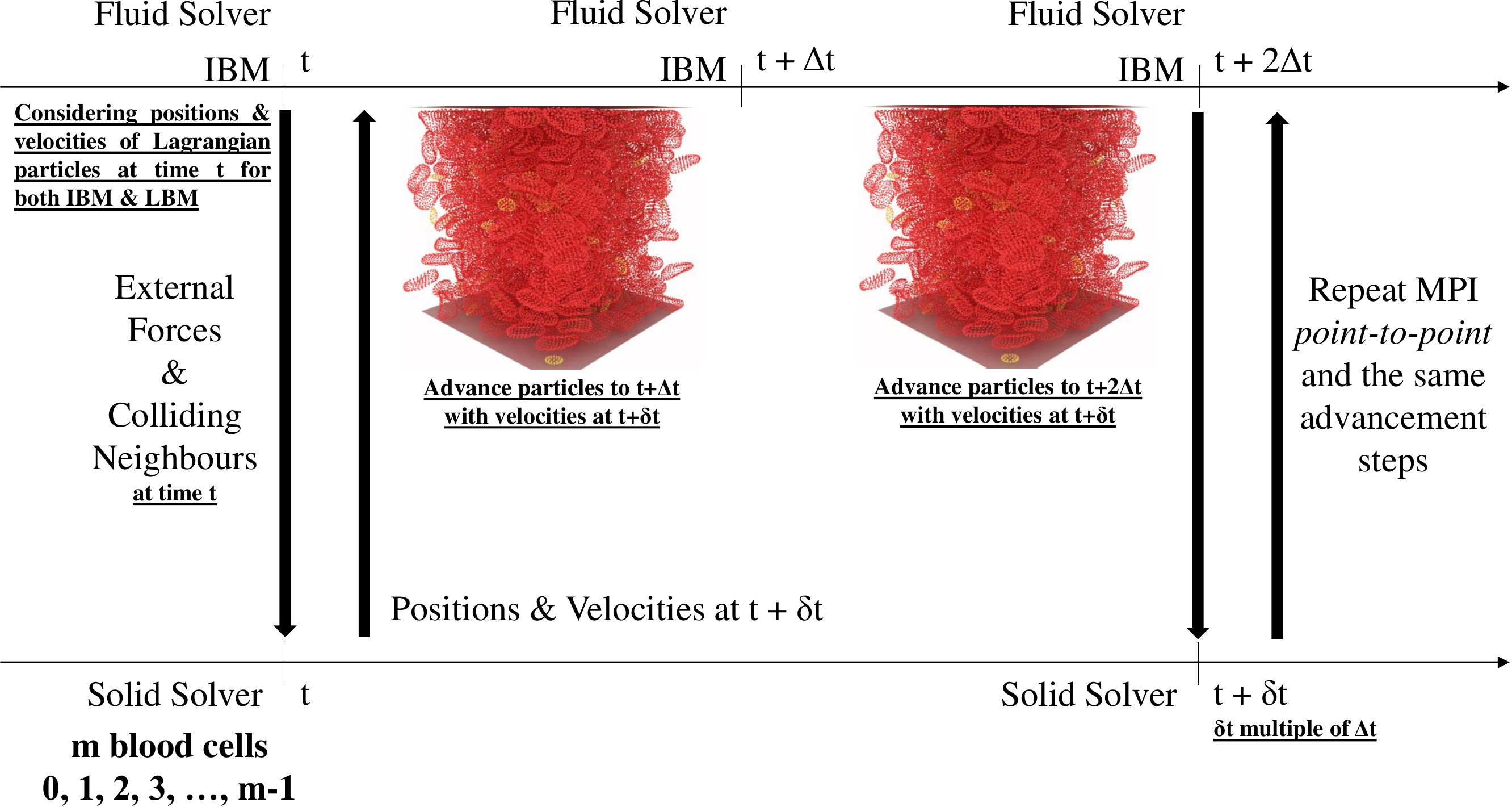}
    \caption{Modular structure of our computational framework. We present the two independent solver streams and the required MPI point-to-point communication for advancing the physical system in time. The decoupling of the solvers leads to a framework that is agnostic to the underlying numerical methods. This diagram remains the same for both the CPU-only version of the framework and the CPU/ GPU implementation (hybrid).}
    \label{fig:Framework_Advancement}
\end{figure}

Load balancing plays a critical role and impacts the efficiency and scalability of HPC applications. For our hybrid CPU/GPU system, three components require special attention. The first is the representation of the fluid domain through a homogeneous grid. The lattice sites are partitioned by Palabos and are distributed to the available MPI tasks (LBM on CPUs). The second component of the system are the plain Lagrangian points that describe the immersed blood cells for the IBM (see Figure \ref{fig:LoadBalancing}, right-hand side image). These points are attached to their immediate fluid cells, and thus their dispatching to the available MPI tasks is aligned with the fluid decomposition (IBM on CPUs). The immersed bodies have a dual nature, i.e. they are seen both as a set of plain Lagrangian points for the IBM but also as a set of augmented Lagrangian particles (connectivity and material properties on top of position and velocity) on the solid solver side, see Figure \ref{fig:LoadBalancing} both images, where both the plain points and the surfaces are represented. This means that the Lagrangian points are duplicated for the IBM and the npFEM modules. Finally, the MPI tasks that are linked with a GPU are responsible for the solid phase. The blood cells are distributed evenly and statically to the available GPUs in a manner that is disconnected from the attribution of the fluid cells and Lagrangian points to the CPUs (npFEM on GPUs). For example, let us assume that MPI task j (see Figure \ref{fig:LoadBalancing}, left-hand side image) handles $m$ blood cells. The blood cell with tag 1 is spatially located in the domain managed by MPI task k. Thus the communication of the external forces, the colliding neighbours and the new state of the body at t+1 is realised through MPI point-to-point communication for all surface vertices of the cell. The same parallel strategy is adopted in the CPU-only version. This strategy may seem counter-intuitive, leading to a substantial communication load, especially compared to an approach in which the structural solver is attributed to nodes dynamically and retains a processor locality with the coupled flow portions. However, the theoretical scalability of our approach is compatible with the massively parallel vision of the project, as the total amount of communicated data scales with the number of blood cells, and it is independent of the number of computational nodes (except in cases with very few nodes). Indeed every surface vertex is involved in exactly two point-to-point exchanges (a fluid-to-solid and a solid-to-fluid exchange). This fact avoids an over-saturation of the network in a situation of weak scaling, provided that the capacity of the network scales with the number of used compute nodes. It further guarantees that our framework can be connected to any structural solver, as the data provided to the structural solver is always fully contained on a compute node. Our approach further ensures a targeted communication strategy, as the data actually needed by the solver can be handpicked. By providing fully decoupled solvers, we favour a generic and modular approach over ad hoc and monolithic solutions. Figure \ref{fig:Framework_Advancement} presents the decoupled structure of our framework, the communication layer and the advancement rules.

\section{Results \& Discussion}  \label{sec:ResAndDisc}
Goal of this section is to prove the capability of our computational framework to efficiently handle problems of varying size. This is done through an exhaustive presentation of performance metrics that are realised at Piz Daint, the flagship system of the Swiss National Supercomputing Centre (CSCS), ranked $6^{th}$ worldwide and $1^{st}$ in Europe according to the list TOP500 (June 2019) \footnote{\url{www.top500.org}}. This supercomputer has 5704 GPU nodes equipped with 12 cores and one NVIDIA GPU each, and 1813 CPU nodes equipped with 36 cores each. A complete presentation of the supercomputer can be found in the supplementary material. Our main focus is the hybrid version of the framework, where the deformable blood cells are resolved on the GPUs, while the blood plasma and the FSI are resolved on the CPUs.

For every case study performed in this work, the flow field has a constant shear rate $100~s^{-1}$, realised with help of a moving top wall and a fixed bottom wall. This low shear rate is chosen in order to reproduce the experiments in Chopard \emph{et al.} \cite{Chopard_PLTs_2017}, and is not due to a computational limitation. Table \ref{tab:RBCs_PLTs_nos} summarises all the different domains, represented through their dimension in $x \times y \times z$ format. The flow direction is parallel to the $z$-axis, the height of the channel spans along the $y$-direction, and the periodic boundaries in $x, z$ directions. Furthermore, the hematocrit of the systems varies between $35\%$ and $45\%$, covering the whole physiological range. The domain is initialised by randomly positioning the blood cells (without avoiding interpenetration) and then executing the computational framework for few thousand steps while the fluid and the FSI solvers are deactivated. This novel cell packing approach is based on the very efficient collision detection offered by Palabos and the unconditional stability of the npFEM solver, which can resolve extreme deformations and interpenetrations. The platelets are simulated as nearly-rigid oblate ellipsoids with a diameter of $3.6~\mu m$, a thickness of $1.1~\mu m$, and a volume of $6.8~fL$, which is an average value for non-activated platelets. The platelet to RBC ratio is $1:5$, and therefore substantially larger than the physiological one ($1:10-20$ \cite{Mehrabadi_2015}). This is a deliberate choice intended to provide more samples for the statistical analysis of the platelet transport. As for the shape of RBCs, the normal biconcave shape is used. A complete list of all the parameters used in this study can be found in the supplementary material. The performance metrics are followed by an analysis on platelet transport. A series of experiments with varying RBC viscoelasticity and channel height present the idiosyncratic behaviour of the platelets and their sensitivity to various flow factors.

\begin{table}[h]
\caption{Numbers of blood cells, RBCs and Platelets (PLTs), for different case studies.}
\label{tab:RBCs_PLTs_nos}
\resizebox{\textwidth}{!}
{
\begin{tabular}{|c|c|c|c|c|c|c|c|c|c|c|}
\hline
\textbf{Computational Domain ($\mu m^3$)} & \multicolumn{2}{c|}{\textbf{50x50x50}} & \multicolumn{2}{c|}{\textbf{50x100x50}} & \multicolumn{2}{c|}{\textbf{50x500x50}} & \multicolumn{2}{c|}{\textbf{50x1000x50}} & \multicolumn{2}{c|}{\textbf{100x1000x100}} \\ \hline
\textbf{PLTs : RBCs = 1:5} & \textbf{RBCs} & \textbf{PLTs} & \textbf{RBCs} & \textbf{PLTs} & \textbf{RBCs} & \textbf{PLTs} & \textbf{RBCs} & \textbf{PLTs} & \textbf{RBCs} & \textbf{PLTs} \\ \hline
\textbf{Hematocrit 35\%} & 476 & 95 & 953 & 190 & 4765 & 953 & 9531 & 1906 & 38126 & 7625 \\ \hline
\textbf{Hematocrit 45\%} & 612 & 122 & 1225 & 245 & 6127 & 1225 & 12255 & 2451 & 49020 & 9804 \\ \hline
\end{tabular}
}
\end{table}

\subsection{Performance Analysis}
Simulations at the spatial scale of millimetre commonly ignore the detailed particulate nature of blood because of the tremendous computational cost, and instead model the effect of the particles through continuum modelling. On the other hand, in publications of state-of-the-art fully resolved whole blood simulations \cite{Vahidkhah_2015}, \cite{BLUMERS2017171}, \cite{Zavodszky2017CellularCells}, \cite{Tan_2018_PALABOS_LAMMPS}, the overall spatial scale of the simulation remains very small, at the order of a few tens of micrometres. The suggested HPC framework is built towards the direction of simulating macroscopic flows, at the order of $mm^3$ of whole blood, while representing the details of the microscopic physics, thus offering users the possibility to address a large range of problems with clinical relevance. In the context of the current scientific goals, the performance metrics of this HPC framework must be considered under the light of weak scaling. Indeed, the purpose of seeking more powerful computational resources is not to improve the resolution or increase the time span of the simulation, but to extend the physical volume of the blood considered in the model.

\begin{figure}[H]
    \centering
    \includegraphics[scale=0.45]{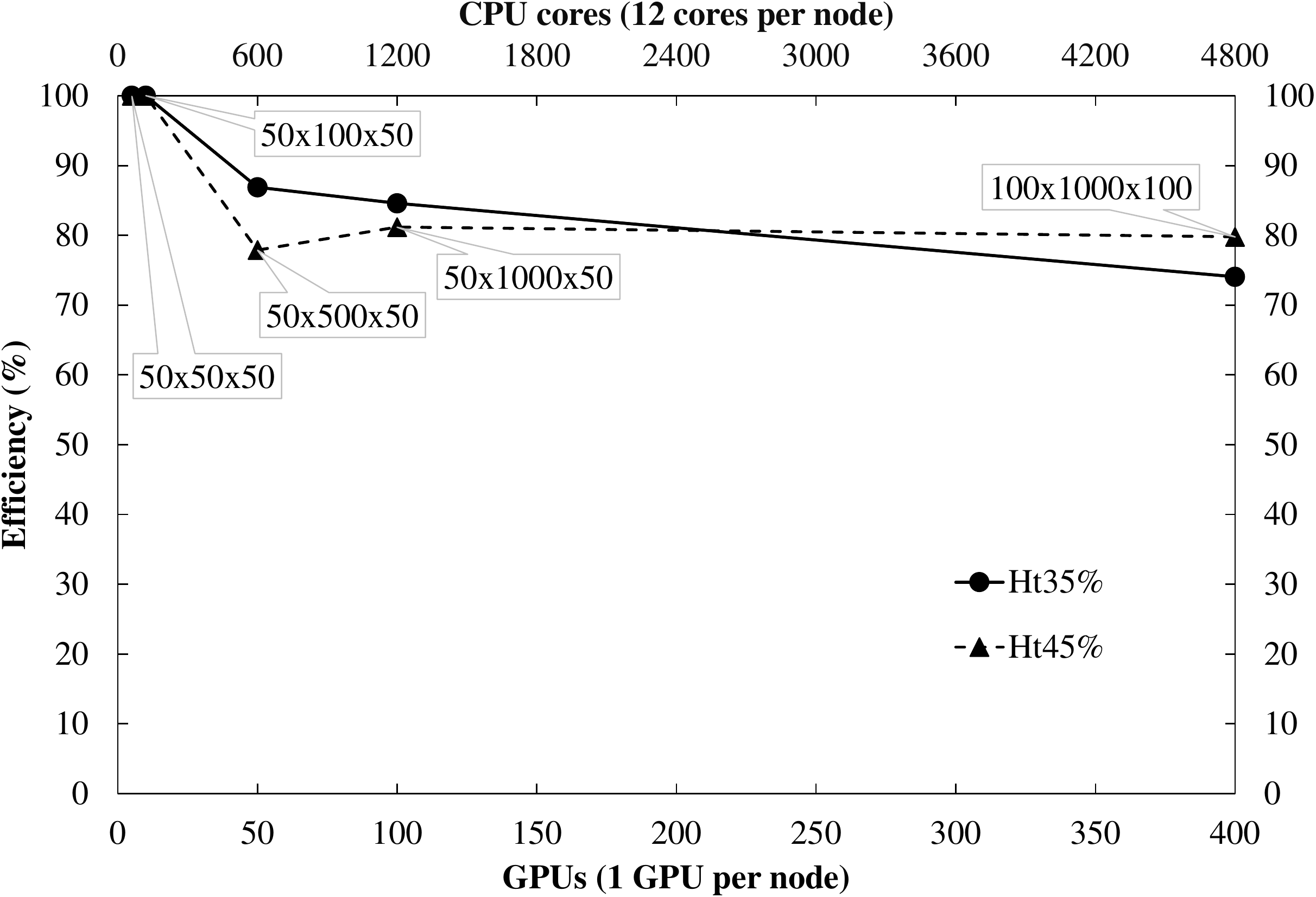}
    \caption{Weak scaling efficiency for various domains at different hematocrit. To better understand the problem sizes consult table \ref{tab:RBCs_PLTs_nos}. The efficiency corresponds to the hybrid version of the code. The GPU nodes in Piz Daint are equipped with 12 cores and one NVIDIA GPU each.}
    \label{fig:WeakScalingHybridEfficiency}
\end{figure}

\begin{figure}[h]
    \centering
    \includegraphics[scale=0.45]{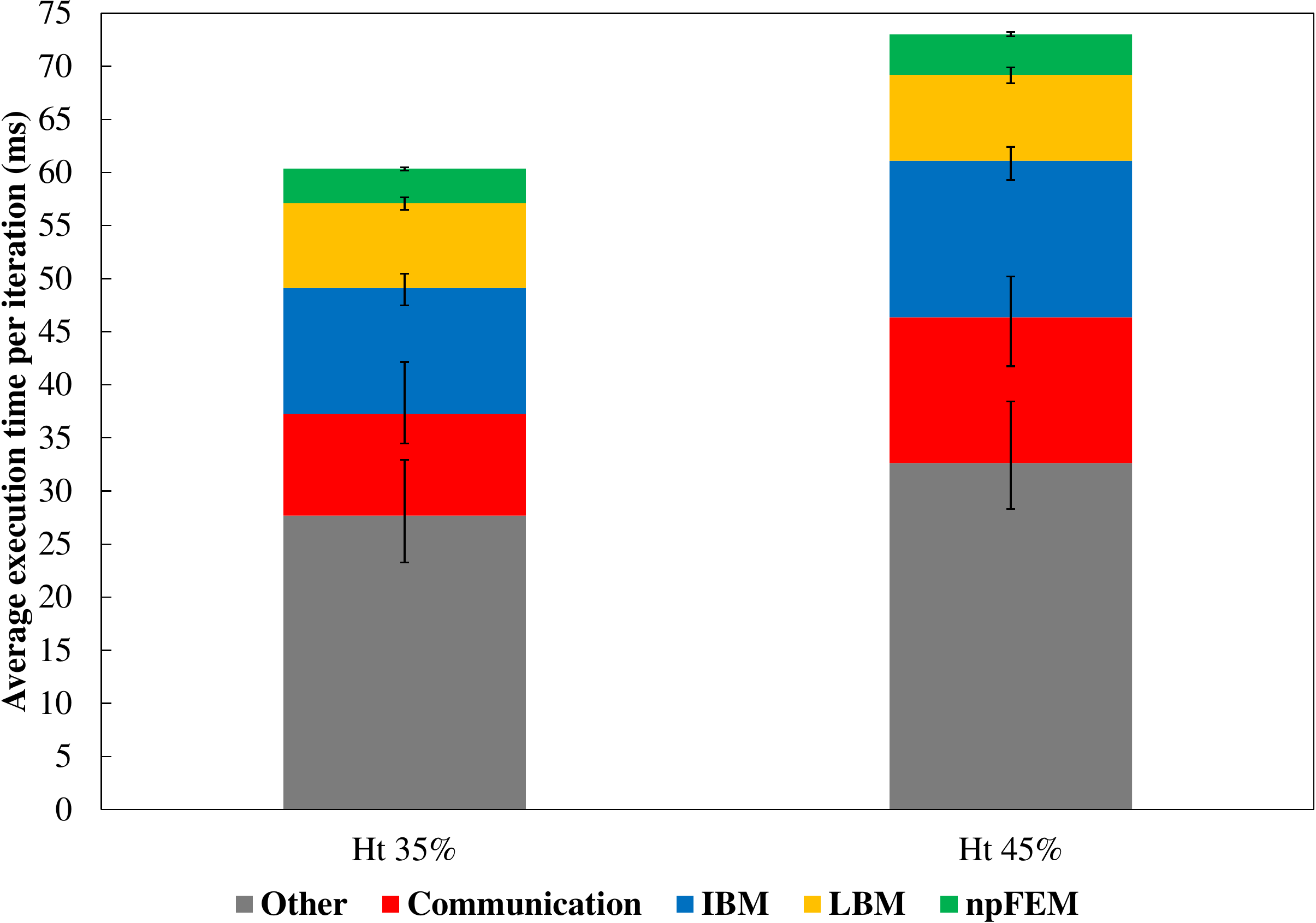}
    \caption{Execution time per iteration for different hematocrit (hybrid version). The ``Other'' contains operations (CPU-only) such as computation of external forces, collisions detection, particles advancement and various book-keeping.}
    \label{fig:StackedHt}
\end{figure}

\begin{figure}[H]
    \centering
    \includegraphics[scale=0.45]{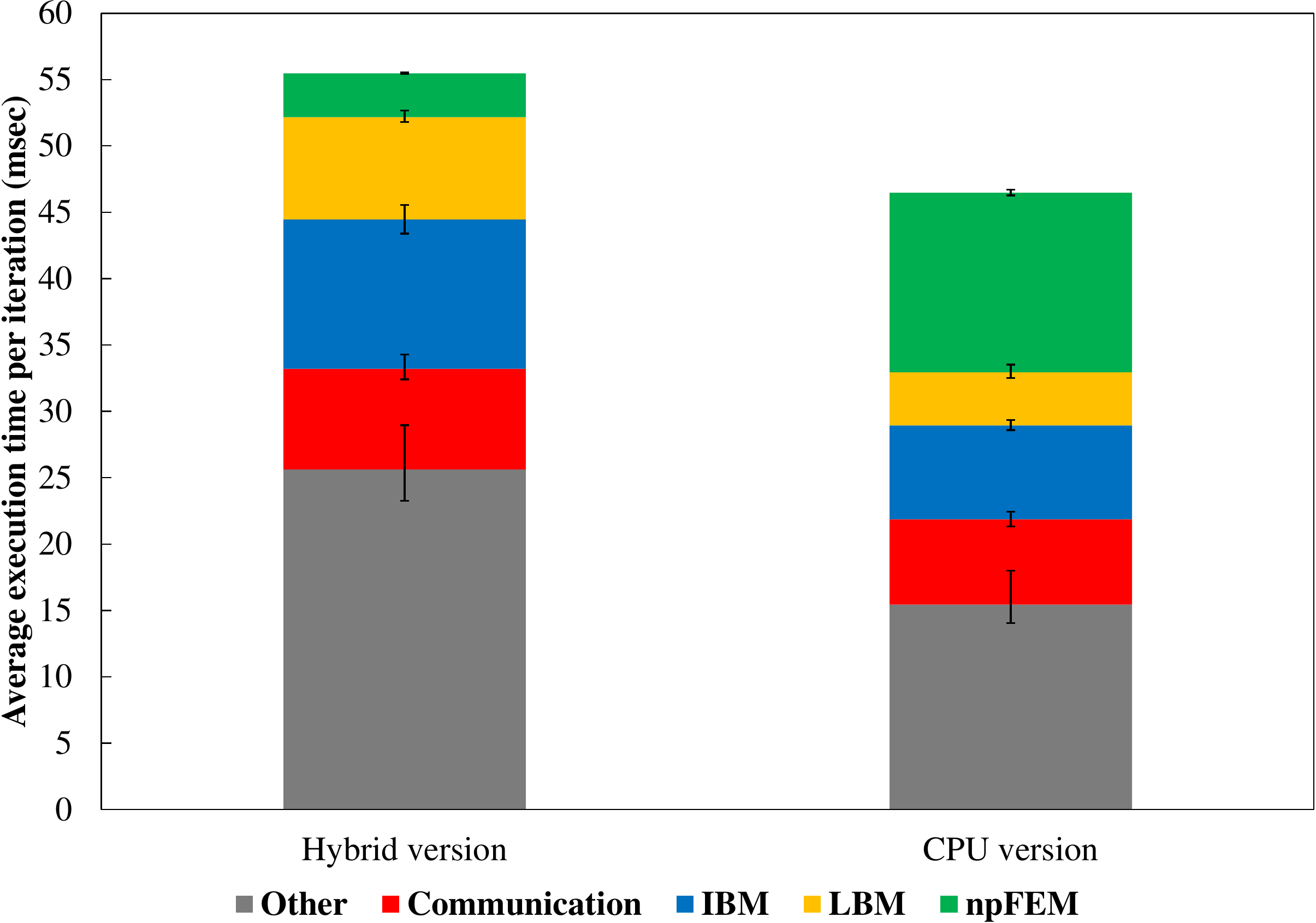}
    \caption{Comparison of execution time per iteration for the Hybrid (CPU/GPU) and CPU-only versions of the framework at 35 \% hematocrit. The GPU nodes have 12 cores and one GPU each, while the CPU nodes have 36 cores each.}
    \label{fig:StackedComp}
\end{figure}

In the weak scaling, the computational load per processor (either CPU or GPU) remains constant. Thus, the problem size increases proportionally with the number of computational hardware units. The reference case study is a 50x50x50 $\mu m^3$ domain, solved on 5 GPU nodes ($N_0$) with reference time noted as $t_{N_0}$. The weak scaling efficiency is given by $\frac{t_{N_0}}{t_N}$, where $t_{N}$ is the time spend in $N$ GPU nodes for a domain $N/N_0$ times larger than the reference one. Figure \ref{fig:WeakScalingHybridEfficiency} presents the efficiency of the proposed computational framework for a problem growth up to $80$ times compared to the reference domain (at 400 GPU nodes). Even if the largest tested domain is still distinctly smaller than $1~mm^3$, the direction of interest (wall-bounded direction) approaches scales of macroscopic extent, and the remaining directions are adequately resolved through periodicity. Our long-term vision for macroscopic flows includes assigning further blood cells per GPU. This on its side requires strong CPU performance to cope with annex preparatory operations (the ``Other'' section on Figures \ref{fig:StackedHt}, \ref{fig:StackedComp}), which might be matched by novel, upcoming supercomputing systems. The code presents good efficiency and its performance does not degrade for higher hematocrit. Other frameworks that are based on a modular approach for the coupling of fluid and solid solvers (\cite{Rahimian_2010}, \cite{Clausen_2010}, \cite{Mountrakis2015ParallelFramework}, \cite{Tan_2018_PALABOS_LAMMPS}) demonstrate an efficiency between $70-80\%$. On the contrary, frameworks that follow the monolithic paradigm \cite{Rossinelli_2015}, deliver a more impressive efficiency, often above $90\%$. Nevertheless, this is a small penalty to be paid for genericity and modularity over ad hoc solutions.

Figure \ref{fig:StackedHt} presents the average execution time per iteration for different hematocrit. The bottom layer of the bars labelled as ``Other'' contains operations (executed on CPUs only) such as computation of external forces, collisions detection, particles advancement and various book-keeping. The ``error'' bars delimit the deviation from the average, where the minimum corresponds to the reference case study and the maximum to the largest case study (see table \ref{tab:RBCs_PLTs_nos}) in the context of weak scaling. A striking observation is that the GPU-ported npFEM solver constitutes only $\sim 6\%$ of the total execution time, especially if compared with other state-of-the-art implementations \cite{Mountrakis2015ParallelFramework}, \cite{Zavodszky2017Hemocell:Library} which report a contribution of the solid solver to over $50\%$ of the overall time. On the other hand, the fluid solver and the FSI account for about $30\%$ of the execution time with a consistent scaling over larger domains and higher hematocrit. The communication is seen to vary around $12-20\%$ of the execution time but does not seem to constitute a bottleneck since it is realised through optimised non-blocking point-to-point communication. The ``Other'' group of operations occupy a large portion of the total time, and this hot-spot reflects the choice of genericity and modularity. A possible conclusion from these observations could be to port the remaining parts of the solver to GPUs given the great performance of the solid solver. It is however debatable if this choice would be optimal, given that modern supercomputers tend to become ``fatter'' in both CPUs and GPUs, as shown in the example of \emph{Summit} with 2 CPUs per node (21 cores each) and 6 NVIDIA Volta GPUs, ranked 1st according to the TOP500 list, June 2019. Thus, the best strategy is to fully exploit the available hardware and not develop one-sided applications. Another counter-argument is that some numerical methods such as the IBM have a large and complex memory footprint that renders them less GPU friendly. An earlier attempt \cite{Beny_2019} to port the whole framework on GPUs could not serve as a justification to move in this direction.

Figure \ref{fig:StackedComp} presents the execution time per iteration and compares the hybrid (CPU/GPU) version with the CPU-only version, including both the measured time averages and the deviations from the average. These results provide further insights into  the weak scaling results up to a domain 50x500x50 $\mu m^3$ at 35\% hematocrit. Main assumption is the one-to-one correspondence of the GPU and CPU nodes of Piz Daint, e.g. solving the computational domain 50x500x50 $\mu m ^3$ in 50 GPU nodes (one GPU and 12 cores each) for the hybrid version and in 50 CPU nodes (36 cores each) for the CPU-only version. The npFEM solver on its own exhibits a speedup of about 4, favouring the GPU implementation. Moreover, in the CPU-only version it is obvious that the solid solver constitutes an overwhelming part of the overall performance, while in the hybrid version the GPU-porting solves this problem in a very efficient manner. More performance analysis results can be found in the supplementary material.

The main challenge of the computational tools for the simulation of the particulate nature of blood is to solve systems with a sufficient number of blood cells ($\gg 1000$) for a physically relevant duration ($\sim$ 1 s) in a reasonable time (less than a week) with the smallest possible amount of computational resources. The proposed computational framework achieves the aforementioned goals and can be compared with other state-of-the-art solvers \cite{Zavodszky2017Hemocell:Library}, \cite{Tan_2018_PALABOS_LAMMPS}. The main novelty is that we are able to achieve this by using high fidelity models for the blood cells which have a richer physical content than simple mass-spring systems. To the best of our knowledge, there is no other computational framework using fully validated FEM-based solid solver that can achieve these target values.

\subsection{Platelet Transport with varying Viscoelasticity of RBCs}
RBC viscoelastic behaviour, a collective term for the contribution of both the membrane and the cytoplasm, is a widely-accepted factor with critical impact on health and disease (pathophysiological conditions). Pathological alterations in RBC deformability have been associated with various diseases \cite{Tomaiuolo2014BiomechanicalMicrofluidics} such as malaria, sickle cell anaemia, diabetes, hereditary disorders, chronic obstructive pulmonary disease, etc. Despite its crucial role, RBC viscoelasticity is overlooked in the majority of the computational tools for cellular flows. Here, we study the effect of different viscoelastic parameters of RBCs on the platelet transport and discriminate each case with the use of the mean square displacement (MSD) in the wall-bounded direction. The parameter that is altered is the $\kappa_{\text{damping}}$ as presented in Kotsalos \emph{et al.} \cite{Kotsalos_JCP_2019_npFEM}. The MSD is defined as $\left \langle (y_i(t) - y_i(t_0))^2 \right \rangle$, with $y_i$ the position of platelet $i$ in the wall-bounded y-direction. The averaging spans either over all the available platelets, i.e. RBC-rich layer (RBC-RL) and Cell Free Layer (CFL), or only over the platelets of the RBC-RL. The flow setup includes a constant shear rate at $100~s^{-1}$, a domain of size $50^3~\mu m^3$ and a hematocrit of $35\%$ (see table \ref{tab:RBCs_PLTs_nos}).

\begin{figure}[h]
    \centering
    \includegraphics[scale=0.45]{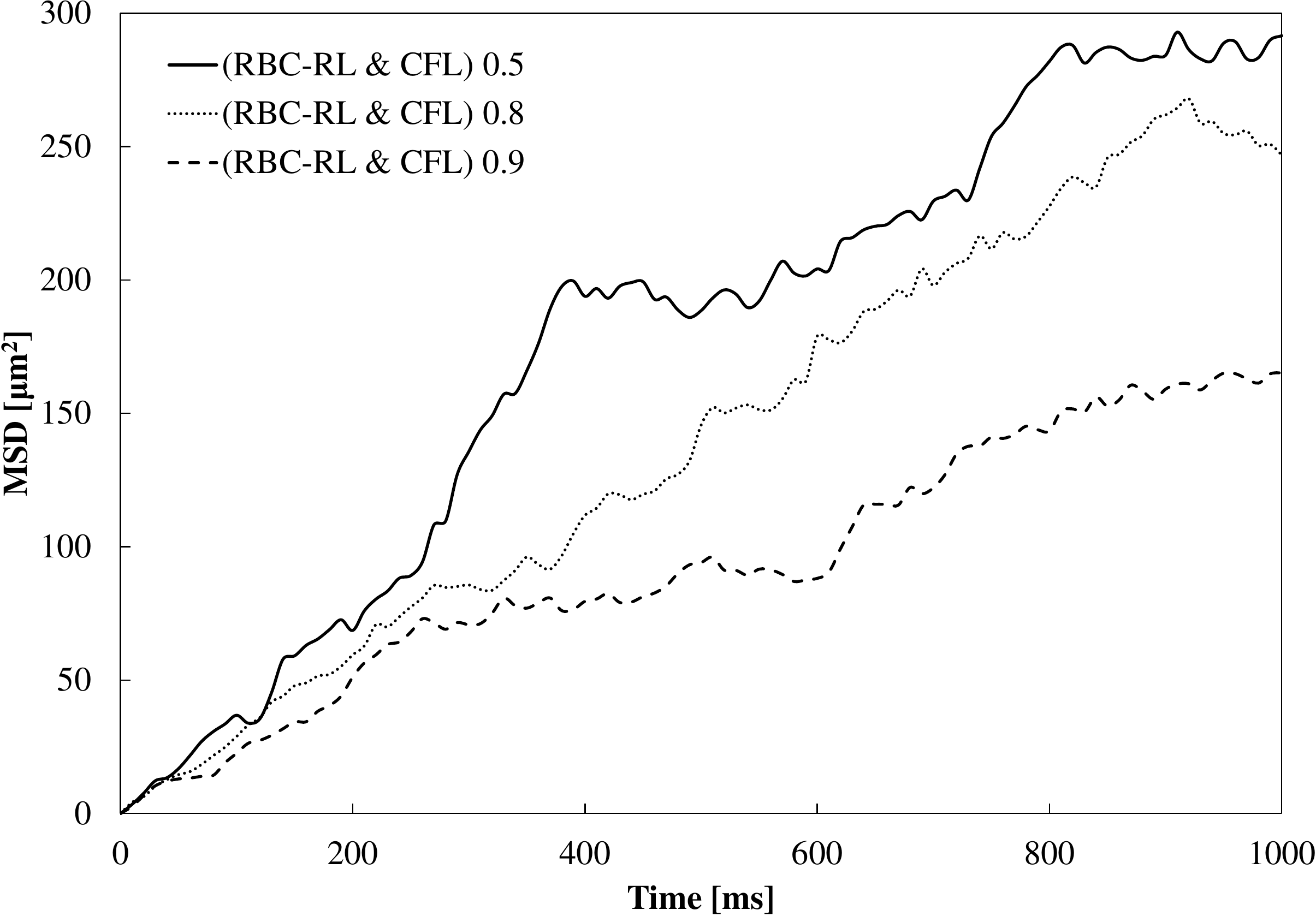}
    \caption{MSD with averaging over all the available platelets. The curves correspond to different values of $\kappa_{\text{damping}} = \{0.5,0.8,0.9\}$.}
    \label{fig:visc_MSD_whole}
\end{figure}

Figure \ref{fig:visc_MSD_whole} presents the MSD over all the platelets of the domain for three different values of the viscoelastic parameter $\kappa_{\text{damping}}$. There is a clear distinction of the less viscous RBCs ($\kappa_{\text{damping}} = 0.5$) to the more viscous ($\kappa_{\text{damping}} = 0.9$) and their projected effect on the platelet transport. The higher the viscoelasticity, the slower the response to the imposed external loads.

\begin{figure}[h]
    \centering
    \includegraphics[scale=0.45]{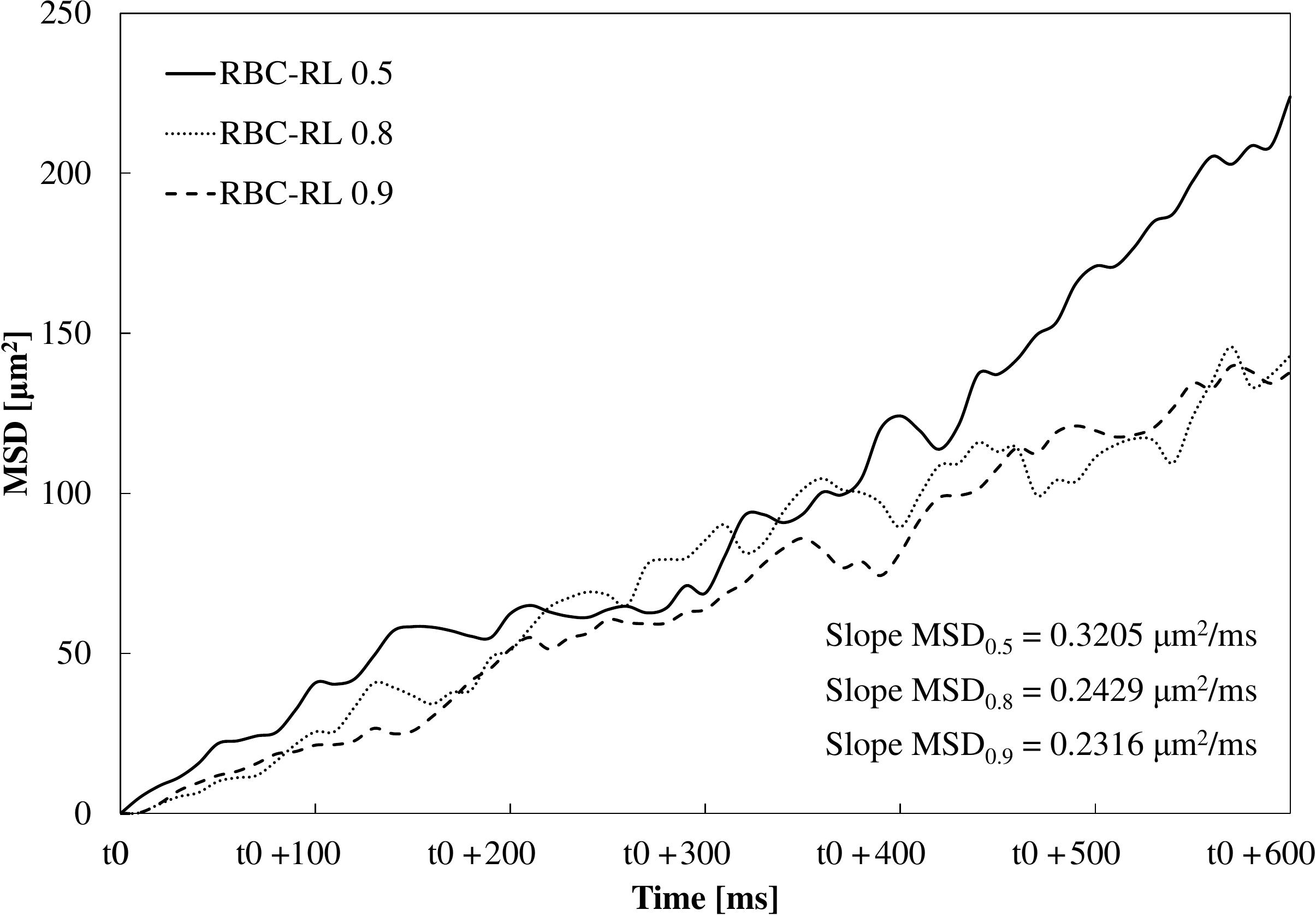}
    \caption{MSD with averaging over the platelets in the RBC-RL. The slopes at the bottom right are obtained from a linear fitting on each curve. The $t_0$ corresponds to 300 ms from the beginning of the simulations. Diffusion coefficients are measured at steady state.}
    \label{fig:visc_MSD_core_only}
\end{figure}

The computation of the diffusion coefficient $D$ of the platelets demands the MSD to be averaged over the RBC-RL, and its slope corresponds to $2D$. Figure \ref{fig:visc_MSD_core_only} presents the MSD and a linear fitting on the curves. The diffusion coefficient in all cases is about $\mathcal{O}\left ( 10^{-10} \right ) ~m^2/s$, in agreement with previously reported values \cite{Zydney_Colton}, \cite{Affeld_2013}, \cite{Vahidkhah_2014}. It is two to three orders of magnitude higher than the Brownian diffusivity, suggesting RBC-augmented diffusion. An interesting observation in conjunction with the study of Kumar and Graham \cite{Kumar_Graham_2011}, \cite{Kumar_Graham_2012} and assuming that the more viscous RBCs are more ``rigid'' (slower response to external forcing), is that the less viscous RBCs $\kappa_{\text{damping}} = 0.5$ lead to higher platelet diffusivity and thus faster concentration towards the walls. The varying diffusivity can be explained from the fact that in heterogeneous collisions the net displacement of the stiff particle (platelet) is substantially larger than that of the floppy particle (RBC) and  the displacement is larger for larger rigidity ratio \cite{Kumar_Graham_2011}. Thus for less viscous RBCs we expect higher displacements of platelets and thus larger diffusion coefficient, as shown in Figure \ref{fig:visc_MSD_core_only}. The platelets that reach the walls tend to stay in the CFL and behave as being trapped in this layer, under any experiment conducted.

\subsection{Platelet Transport for Larger Geometries}
\begin{figure}[h]
    \centering
    \includegraphics[scale=0.45]{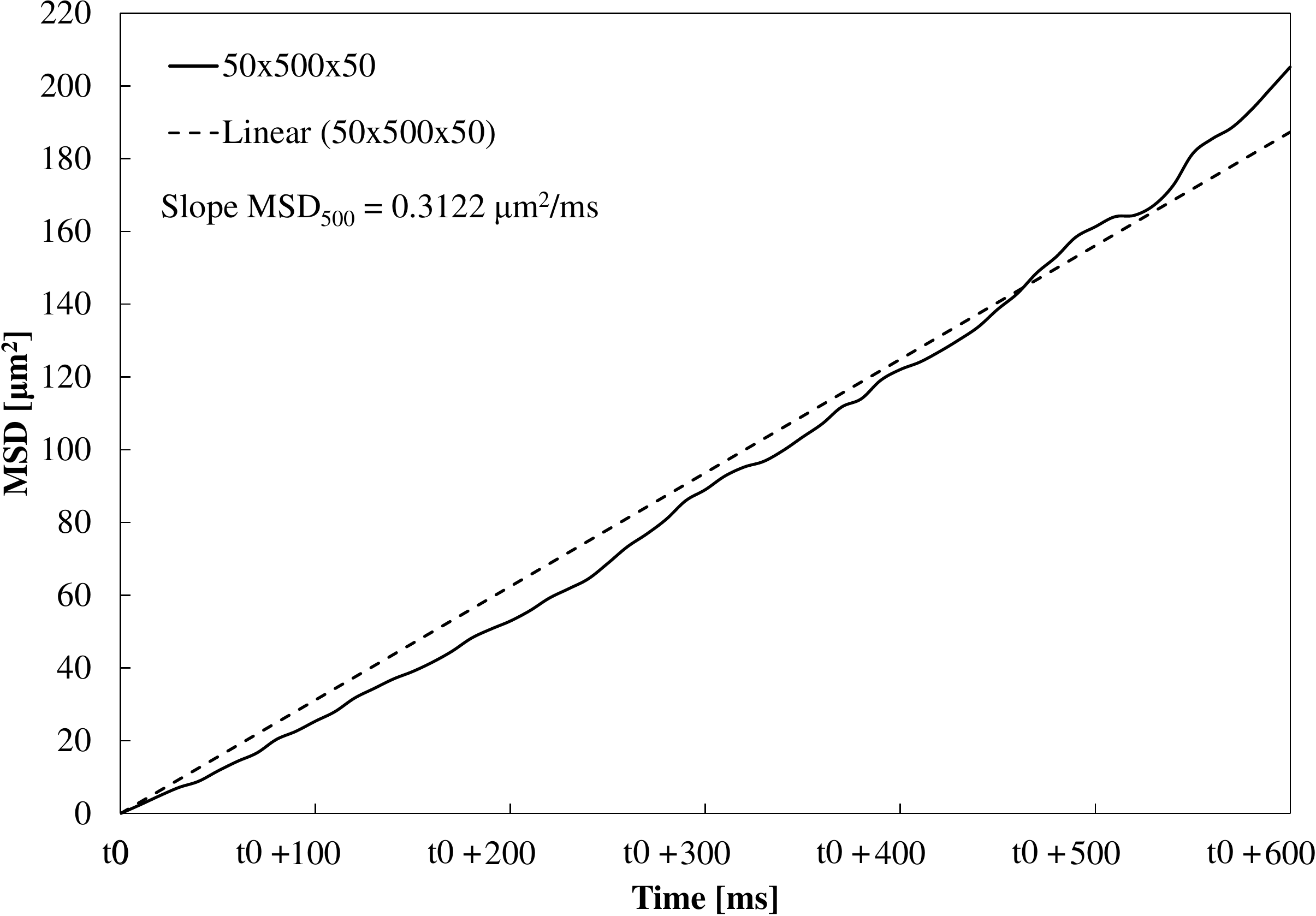}
    \caption{Mean square displacement measured in the RBC-rich layer. The $t_0$ corresponds to 300 ms from the beginning of the simulations. The linear fitting on the MSD gives a slope that is equal to 2D, where D is the diffusion coefficient of the platelets.}
    \label{fig:sizePartial_MSD_core_only}
\end{figure}

\begin{figure}[h]
    \centering
    \includegraphics[scale=0.7]{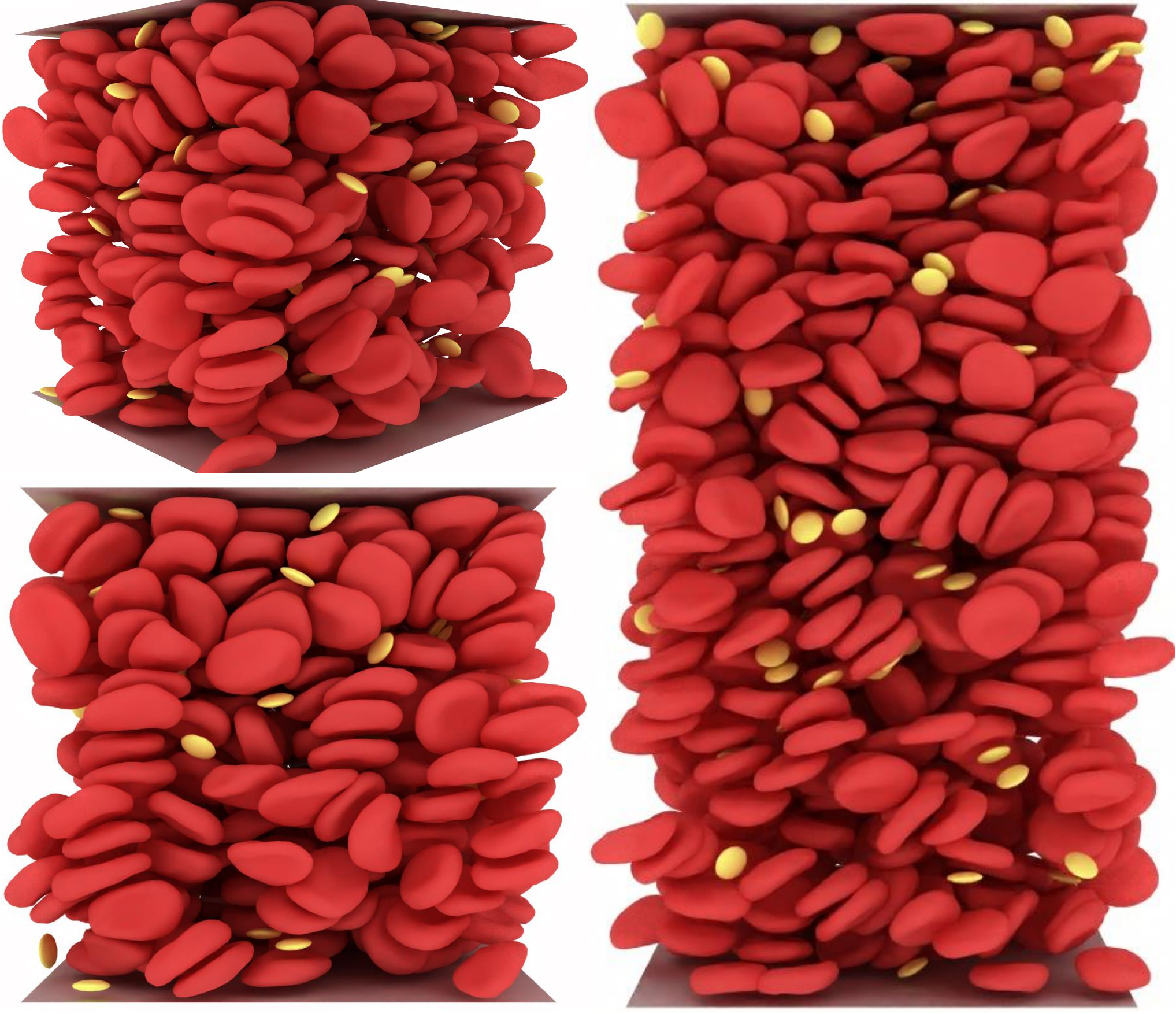}
    \caption{Shear flow generated by our computational framework for fully resolved blood flow simulations. The left image shows two different viewpoints of the $50^3~\mu m^3$ domain at 35\% hematocrit. The right image depicts a domain 50x100x50 $\mu m^3$ at 35 \% hematocrit.}
    \label{fig:Sims}
\end{figure}

Most studies are bounded by domains of few micrometers and low hematocrit due to the high computational cost. Nevertheless, interesting phenomena can amplify as sizes increase \cite{Mountrakis_whereDoPLTsGo}. Given our HPC-capable framework, we are interested on quantifying the diffusivity of the platelets as the channel height varies. Here, a flow field with constant shear rate $100~s^{-1}$ and $35\%$ hematocrit is considered. The wall-bounded direction takes three different sizes $H=\left \{ 50, 100, 500 \right \} ~\mu m$, while the other, periodic directions remain at $50~\mu m$ (see table \ref{tab:RBCs_PLTs_nos}). The dimensionless numbers that describe the dynamics of the problem are the capsule Reynolds number $Re_{capsule} = \frac{\dot{\gamma} r^2}{\nu}$ with $\dot{\gamma}$ the shear rate, $r$ the characteristic length of the capsule, and the capillary number $Ca = \frac{\mu \dot{\gamma} r}{B_{Skalak}}$ with $\mu$ the dynamic viscosity of blood plasma and $B_{Skalak}$ the membrane shear modulus (see supplementary material and \cite{Kotsalos_JCP_2019_npFEM}). Figure \ref{fig:sizePartial_MSD_core_only} shows the mean square displacement in the RBC-RL for the largest case study along with the linear fitting. The diffusion coefficients for all the different experiments are about two to three orders of magnitude higher than the Brownian diffusivity \cite{Vahidkhah_2014}, \cite{Zydney_Colton}. Figure \ref{fig:Sims} summarises the simulations conducted for the varying channel case study.

\section{Conclusions}
In this study, we provided a computational framework for digital blood. The full resolution of the particulate nature of blood is a challenging venture, especially when it is compiled into a framework that is based on generality, modularity, performance without compromising robustness and accuracy. The individual numerical techniques used for the simulation of blood constituents (LBM for the fluid and FEM for the solid phase) are characterised by their high fidelity for capturing physical phenomena, and their coupling has shown to sufficiently resolve the complex interaction between the blood cells.

This kind of computational tool complements the toolset for a digital lab. More precisely, the present project complements another research activity based on a coarse-grained approximation of blood using stochastic methods and random walks. The fully resolved models, apart from providing in-depth investigations on various case studies, are used to fine-tune the coarse-grained models, e.g. providing diffusion coefficients of various particles, thus constituting a critical component in this integrative approach towards digital blood. Our future scientific endeavours will be moving to this multi-scale direction as recently depicted in \cite{Randles_2019}.

\section*{Acknowledgements}
We acknowledge support from the PASC project 2017-2020, Virtual Physiological Blood: an HPC framework for blood flow simulations in vasculature and in medical devices.

\section*{Funding}
This project has received funding from the European Union’s Horizon 2020 research and innovation programme under grant agreement No 823712 (CompBioMed2 project).


\clearpage
\bibliography{References}

\begin{thebibliography}{46}
\providecommand{\natexlab}[1]{#1}
\providecommand{\url}[1]{\texttt{#1}}
\providecommand{\href}[2]{#2}
\providecommand{\path}[1]{#1}
\providecommand{\eprint}[1]{\href{http://arxiv.org/abs/#1}{\path{#1}}}
\providecommand{\DOIprefix}{doi:}
\providecommand{\ArXivprefix}{arXiv:}
\providecommand{\URLprefix}{URL: }
\providecommand{\Pubmedprefix}{pmid:}
\providecommand{\doi}[1]{\href{http://dx.doi.org/#1}{\path{#1}}}
\providecommand{\Pubmed}[1]{\href{pmid:#1}{\path{#1}}}
\providecommand{\BIBand}{and}
\providecommand{\bibinfo}[2]{#2}
\ifx\xfnm\undefined \def\xfnm[#1]{\unskip,\space#1}\fi
\bibitem[{Tomaiuolo(2014)}]{Tomaiuolo2014BiomechanicalMicrofluidics}
\bibinfo{author}{Tomaiuolo\xfnm[ G.]}.
\newblock \bibinfo{title}{Biomechanical properties of red blood cells in health
  and disease towards microfluidics}.
\newblock \bibinfo{journal}{Biomicrofluidics}
  \bibinfo{year}{2014};\bibinfo{volume}{8}(\bibinfo{number}{5}):\bibinfo{pages}{51501}.
\newblock \DOIprefix\doi{10.1063/1.4895755}.
\bibitem[{Freund(2014)}]{Freund_2014}
\bibinfo{author}{Freund\xfnm[ J.B.]}.
\newblock \bibinfo{title}{Numerical simulation of flowing blood cells}.
\newblock \bibinfo{journal}{Annual Review of Fluid Mechanics}
  \bibinfo{year}{2014};\bibinfo{volume}{46}(\bibinfo{number}{1}):\bibinfo{pages}{67--95}.
\newblock \DOIprefix\doi{10.1146/annurev-fluid-010313-141349}.
\bibitem[{Z{\'{a}}vodszky et~al.(2017)Z{\'{a}}vodszky, van Rooij, Azizi and
  Hoekstra}]{Zavodszky2017CellularCells}
\bibinfo{author}{Z{\'{a}}vodszky\xfnm[ G.]}, \bibinfo{author}{van Rooij\xfnm[
  B.]}, \bibinfo{author}{Azizi\xfnm[ V.]}, \bibinfo{author}{Hoekstra\xfnm[
  A.]}.
\newblock \bibinfo{title}{Cellular level in-silico modeling of blood rheology
  with an improved material model for red blood cells}.
\newblock \bibinfo{journal}{Frontiers in Physiology}
  \bibinfo{year}{2017};\bibinfo{volume}{8}:\bibinfo{pages}{1--14}.
\newblock \DOIprefix\doi{10.3389/fphys.2017.00563}.
\bibitem[{Fogelson and Neeves(2015)}]{Fogelson_Review_2015}
\bibinfo{author}{Fogelson\xfnm[ A.L.]}, \bibinfo{author}{Neeves\xfnm[ K.B.]}.
\newblock \bibinfo{title}{Fluid mechanics of blood clot formation}.
\newblock \bibinfo{journal}{Annual Review of Fluid Mechanics}
  \bibinfo{year}{2015};\bibinfo{volume}{47}(\bibinfo{number}{1}):\bibinfo{pages}{377--403}.
\newblock \DOIprefix\doi{10.1146/annurev-fluid-010814-014513}.
\bibitem[{Vahidkhah et~al.(2014)Vahidkhah, Diamond and Bagchi}]{Vahidkhah_2014}
\bibinfo{author}{Vahidkhah\xfnm[ K.]}, \bibinfo{author}{Diamond\xfnm[ S.]},
  \bibinfo{author}{Bagchi\xfnm[ P.]}.
\newblock \bibinfo{title}{Platelet dynamics in three-dimensional simulation of
  whole blood}.
\newblock \bibinfo{journal}{Biophysical Journal}
  \bibinfo{year}{2014};\bibinfo{volume}{106}(\bibinfo{number}{11}):\bibinfo{pages}{2529
  -- 2540}.
\newblock \DOIprefix\doi{10.1016/j.bpj.2014.04.028}.
\bibitem[{Mehrabadi et~al.(2015)Mehrabadi, Ku and Aidun}]{Mehrabadi_2015}
\bibinfo{author}{Mehrabadi\xfnm[ M.]}, \bibinfo{author}{Ku\xfnm[ D.N.]},
  \bibinfo{author}{Aidun\xfnm[ C.K.]}.
\newblock \bibinfo{title}{A continuum model for platelet transport in flowing
  blood based on direct numerical simulations of cellular blood flow}.
\newblock \bibinfo{journal}{Annals of Biomedical Engineering}
  \bibinfo{year}{2015};\bibinfo{volume}{43}(\bibinfo{number}{6}):\bibinfo{pages}{1410--1421}.
\newblock \DOIprefix\doi{10.1007/s10439-014-1168-4}.
\bibitem[{Zhao and Shaqfeh(2011)}]{Zhao_PLTs_2011}
\bibinfo{author}{Zhao\xfnm[ H.]}, \bibinfo{author}{Shaqfeh\xfnm[ E.S.G.]}.
\newblock \bibinfo{title}{Shear-induced platelet margination in a
  microchannel}.
\newblock \bibinfo{journal}{Phys Rev E}
  \bibinfo{year}{2011};\bibinfo{volume}{83}:\bibinfo{pages}{061924}.
\newblock \DOIprefix\doi{10.1103/PhysRevE.83.061924}.
\bibitem[{Fedosov et~al.(2011)Fedosov, Caswell, Suresh and
  Karniadakis}]{Fedosov_2011a}
\bibinfo{author}{Fedosov\xfnm[ D.A.]}, \bibinfo{author}{Caswell\xfnm[ B.]},
  \bibinfo{author}{Suresh\xfnm[ S.]}, \bibinfo{author}{Karniadakis\xfnm[
  G.E.]}.
\newblock \bibinfo{title}{Quantifying the biophysical characteristics of
  plasmodium-falciparum-parasitized red blood cells in microcirculation}.
\newblock \bibinfo{journal}{Proceedings of the National Academy of Sciences}
  \bibinfo{year}{2011};\bibinfo{volume}{108}(\bibinfo{number}{1}):\bibinfo{pages}{35--39}.
\newblock \DOIprefix\doi{10.1073/pnas.1009492108}.
\bibitem[{Li et~al.(2017)Li, Li, Chang, Lykotrafitis and
  Em~Karniadakis}]{Li_2017}
\bibinfo{author}{Li\xfnm[ X.]}, \bibinfo{author}{Li\xfnm[ H.]},
  \bibinfo{author}{Chang\xfnm[ H.Y.]}, \bibinfo{author}{Lykotrafitis\xfnm[
  G.]}, \bibinfo{author}{Em~Karniadakis\xfnm[ G.]}.
\newblock \bibinfo{title}{{Computational Biomechanics of Human Red Blood Cells
  in Hematological Disorders}}.
\newblock \bibinfo{journal}{Journal of Biomechanical Engineering}
  \bibinfo{year}{2017};\bibinfo{volume}{139}(\bibinfo{number}{2}).
\newblock \DOIprefix\doi{10.1115/1.4035120}.
\bibitem[{Chang et~al.(2018)Chang, Yazdani, Li, Douglas, Mantzoros and
  Karniadakis}]{Chang_Karniadakis_2018}
\bibinfo{author}{Chang\xfnm[ H.Y.]}, \bibinfo{author}{Yazdani\xfnm[ A.]},
  \bibinfo{author}{Li\xfnm[ X.]}, \bibinfo{author}{Douglas\xfnm[ K.A.]},
  \bibinfo{author}{Mantzoros\xfnm[ C.S.]}, \bibinfo{author}{Karniadakis\xfnm[
  G.E.]}.
\newblock \bibinfo{title}{Quantifying platelet margination in diabetic
  blood flow}.
\newblock \bibinfo{journal}{Biophysical Journal}
  \bibinfo{year}{2018};\bibinfo{volume}{115}(\bibinfo{number}{7}):\bibinfo{pages}{1371
  -- 1382}.
\newblock \DOIprefix\doi{10.1016/j.bpj.2018.08.031}.
\bibitem[{{Rossinelli} et~al.(2015){Rossinelli}, {Tang}, {Lykov}, {Alexeev},
  {Bernaschi}, {Hadjidoukas} et~al.}]{Rossinelli_2015}
\bibinfo{author}{{Rossinelli}\xfnm[ D.]}, \bibinfo{author}{{Tang}\xfnm[ Y.]},
  \bibinfo{author}{{Lykov}\xfnm[ K.]}, \bibinfo{author}{{Alexeev}\xfnm[ D.]},
  \bibinfo{author}{{Bernaschi}\xfnm[ M.]}, \bibinfo{author}{{Hadjidoukas}\xfnm[
  P.]}, et~al.
\newblock \bibinfo{title}{The in-silico lab-on-a-chip: petascale and
  high-throughput simulations of microfluidics at cell resolution}.
\newblock In: \bibinfo{booktitle}{SC '15: Proceedings of the International
  Conference for High Performance Computing, Networking, Storage and Analysis}.
  \bibinfo{year}{2015}, p. \bibinfo{pages}{1--12}.
\newblock \DOIprefix\doi{10.1145/2807591.2807677}.
\bibitem[{Krüger et~al.(2014)Krüger, Holmes and Coveney}]{Kruger_2014}
\bibinfo{author}{Krüger\xfnm[ T.]}, \bibinfo{author}{Holmes\xfnm[ D.]},
  \bibinfo{author}{Coveney\xfnm[ P.V.]}.
\newblock \bibinfo{title}{Deformability-based red blood cell separation in
  deterministic lateral displacement devices—a simulation study}.
\newblock \bibinfo{journal}{Biomicrofluidics}
  \bibinfo{year}{2014};\bibinfo{volume}{8}(\bibinfo{number}{5}):\bibinfo{pages}{054114}.
\newblock \DOIprefix\doi{10.1063/1.4897913}.
\bibitem[{Sen~Gupta(2016)}]{Gupta_2016}
\bibinfo{author}{Sen~Gupta\xfnm[ A.]}.
\newblock \bibinfo{title}{Role of particle size, shape, and stiffness in design
  of intravascular drug delivery systems: insights from computations,
  experiments, and nature}.
\newblock \bibinfo{journal}{Wiley Interdisciplinary Reviews: Nanomedicine and
  Nanobiotechnology}
  \bibinfo{year}{2016};\bibinfo{volume}{8}(\bibinfo{number}{2}):\bibinfo{pages}{255--270}.
\newblock \DOIprefix\doi{10.1002/wnan.1362}.
\bibitem[{Vahidkhah and Bagchi(2015)}]{Vahidkhah_2015}
\bibinfo{author}{Vahidkhah\xfnm[ K.]}, \bibinfo{author}{Bagchi\xfnm[ P.]}.
\newblock \bibinfo{title}{Microparticle shape effects on margination{,}
  near-wall dynamics and adhesion in a three-dimensional simulation of red
  blood cell suspension}.
\newblock \bibinfo{journal}{Soft Matter}
  \bibinfo{year}{2015};\bibinfo{volume}{11}:\bibinfo{pages}{2097--2109}.
\newblock \DOIprefix\doi{10.1039/C4SM02686A}.
\bibitem[{Kotsalos et~al.(2019)Kotsalos, Latt and
  Chopard}]{Kotsalos_JCP_2019_npFEM}
\bibinfo{author}{Kotsalos\xfnm[ C.]}, \bibinfo{author}{Latt\xfnm[ J.]},
  \bibinfo{author}{Chopard\xfnm[ B.]}.
\newblock \bibinfo{title}{Bridging the computational gap between mesoscopic and
  continuum modeling of red blood cells for fully resolved blood flow}.
\newblock \bibinfo{journal}{Journal of Computational Physics}
  \bibinfo{year}{2019};\bibinfo{volume}{398}.
\newblock \DOIprefix\doi{10.1016/j.jcp.2019.108905}.
\bibitem[{{Rahimian} et~al.(2010){Rahimian}, {Lashuk}, {Veerapaneni},
  {Chandramowlishwaran}, {Malhotra}, {Moon} et~al.}]{Rahimian_2010}
\bibinfo{author}{{Rahimian}\xfnm[ A.]}, \bibinfo{author}{{Lashuk}\xfnm[ I.]},
  \bibinfo{author}{{Veerapaneni}\xfnm[ S.]},
  \bibinfo{author}{{Chandramowlishwaran}\xfnm[ A.]},
  \bibinfo{author}{{Malhotra}\xfnm[ D.]}, \bibinfo{author}{{Moon}\xfnm[ L.]},
  et~al.
\newblock \bibinfo{title}{Petascale direct numerical simulation of blood flow
  on 200k cores and heterogeneous architectures}.
\newblock In: \bibinfo{booktitle}{SC '10: Proceedings of the 2010 ACM/IEEE
  International Conference for High Performance Computing, Networking, Storage
  and Analysis}. \bibinfo{year}{2010}, p. \bibinfo{pages}{1--11}.
\newblock \DOIprefix\doi{10.1109/SC.2010.42}.
\bibitem[{{Peters} et~al.(2010){Peters}, {Melchionna}, {Kaxiras}, {Lätt},
  {Sircar}, {Bernaschi} et~al.}]{Peters_2010}
\bibinfo{author}{{Peters}\xfnm[ A.]}, \bibinfo{author}{{Melchionna}\xfnm[ S.]},
  \bibinfo{author}{{Kaxiras}\xfnm[ E.]}, \bibinfo{author}{{Lätt}\xfnm[ J.]},
  \bibinfo{author}{{Sircar}\xfnm[ J.]}, \bibinfo{author}{{Bernaschi}\xfnm[
  M.]}, et~al.
\newblock \bibinfo{title}{Multiscale simulation of cardiovascular flows on the
  ibm bluegene/p: Full heart-circulation system at red-blood cell resolution}.
\newblock In: \bibinfo{booktitle}{SC '10: Proceedings of the 2010 ACM/IEEE
  International Conference for High Performance Computing, Networking, Storage
  and Analysis}. \bibinfo{year}{2010}, p. \bibinfo{pages}{1--10}.
\newblock \DOIprefix\doi{10.1109/SC.2010.33}.
\bibitem[{Bernaschi et~al.(2011)Bernaschi, Bisson, Endo, Matsuoka, Fatica and
  Melchionna}]{Bernaschi_2011}
\bibinfo{author}{Bernaschi\xfnm[ M.]}, \bibinfo{author}{Bisson\xfnm[ M.]},
  \bibinfo{author}{Endo\xfnm[ T.]}, \bibinfo{author}{Matsuoka\xfnm[ S.]},
  \bibinfo{author}{Fatica\xfnm[ M.]}, \bibinfo{author}{Melchionna\xfnm[ S.]}.
\newblock \bibinfo{title}{Petaflop biofluidics simulations on a two
  million-core system}.
\newblock In: \bibinfo{booktitle}{Proceedings of 2011 International Conference
  for High Performance Computing, Networking, Storage and Analysis}. SC '11;
  \bibinfo{address}{New York, NY, USA}: \bibinfo{publisher}{ACM};
  \bibinfo{year}{2011}, p. \bibinfo{pages}{4:1--4:12}.
\newblock \DOIprefix\doi{10.1145/2063384.2063389}.
\bibitem[{Xu et~al.(2013)Xu, Kaliviotis, Munjiza, Avital, Ji and
  Williams}]{Xu_2013_RBCs}
\bibinfo{author}{Xu\xfnm[ D.]}, \bibinfo{author}{Kaliviotis\xfnm[ E.]},
  \bibinfo{author}{Munjiza\xfnm[ A.]}, \bibinfo{author}{Avital\xfnm[ E.]},
  \bibinfo{author}{Ji\xfnm[ C.]}, \bibinfo{author}{Williams\xfnm[ J.]}.
\newblock \bibinfo{title}{Large scale simulation of red blood cell aggregation
  in shear flows}.
\newblock \bibinfo{journal}{Journal of Biomechanics}
  \bibinfo{year}{2013};\bibinfo{volume}{46}(\bibinfo{number}{11}):\bibinfo{pages}{1810
  -- 1817}.
\newblock \DOIprefix\doi{10.1016/j.jbiomech.2013.05.010}.
\bibitem[{Xu et~al.(2017)Xu, Ji, Avital, Kaliviotis, Munjiza and
  Williams}]{Xu_2017_RBCs}
\bibinfo{author}{Xu\xfnm[ D.]}, \bibinfo{author}{Ji\xfnm[ C.]},
  \bibinfo{author}{Avital\xfnm[ E.]}, \bibinfo{author}{Kaliviotis\xfnm[ E.]},
  \bibinfo{author}{Munjiza\xfnm[ A.]}, \bibinfo{author}{Williams\xfnm[ J.]}.
\newblock \bibinfo{title}{{An Investigation on the Aggregation and Rheodynamics
  of Human Red Blood Cells Using High Performance Computations}}.
\newblock \bibinfo{journal}{Scientifica}
  \bibinfo{year}{2017};\DOIprefix\doi{10.1155/2017/6524156}.
\bibitem[{Latt et~al.(2019)Latt, Malaspinas, Kontaxakis, Parmigiani, Lagrava,
  Brogi et~al.}]{PalabosArticle}
\bibinfo{author}{Latt\xfnm[ J.]}, \bibinfo{author}{Malaspinas\xfnm[ O.]},
  \bibinfo{author}{Kontaxakis\xfnm[ D.]}, \bibinfo{author}{Parmigiani\xfnm[
  A.]}, \bibinfo{author}{Lagrava\xfnm[ D.]}, \bibinfo{author}{Brogi\xfnm[ F.]},
  et~al.
\newblock \bibinfo{title}{Palabos: Parallel lattice boltzmann solver}
  \bibinfo{year}{2019};\DOIprefix\doi{10.13140/RG.2.2.20836.94086}.
\bibitem[{Pal(2019)}]{PalabosWebsite}
\bibinfo{title}{Palabos}.
\newblock \bibinfo{howpublished}{\url{https://palabos.unige.ch/}};
  \bibinfo{year}{2019}.
\bibitem[{Dupin et~al.(2007)Dupin, Halliday, Care, Alboul and
  Munn}]{Dupin2007ModelingDimensions}
\bibinfo{author}{Dupin\xfnm[ M.M.]}, \bibinfo{author}{Halliday\xfnm[ I.]},
  \bibinfo{author}{Care\xfnm[ C.M.]}, \bibinfo{author}{Alboul\xfnm[ L.]},
  \bibinfo{author}{Munn\xfnm[ L.L.]}.
\newblock \bibinfo{title}{Modeling the flow of dense suspensions of deformable
  particles in three dimensions}.
\newblock \bibinfo{journal}{Physical review E}
  \bibinfo{year}{2007};\bibinfo{volume}{75}(\bibinfo{number}{6 Pt 2}).
\newblock \DOIprefix\doi{10.1103/PhysRevE.75.066707}.
\bibitem[{Fedosov et~al.(2010)Fedosov, Caswell and
  Karniadakis}]{Fedosov2010ARheologydynamics}
\bibinfo{author}{Fedosov\xfnm[ D.A.]}, \bibinfo{author}{Caswell\xfnm[ B.]},
  \bibinfo{author}{Karniadakis\xfnm[ G.E.]}.
\newblock \bibinfo{title}{A multiscale red blood cell model with accurate
  mechanics, rheology,dynamics}.
\newblock \bibinfo{journal}{Biophysical Journal}
  \bibinfo{year}{2010};\bibinfo{volume}{98}(\bibinfo{number}{10}):\bibinfo{pages}{2215--2225}.
\newblock \DOIprefix\doi{10.1016/j.bpj.2010.02.002}.
\bibitem[{Reasor et~al.(2011)Reasor, Clausen and
  Aidun}]{Reasor2011CouplingFlow}
\bibinfo{author}{Reasor\xfnm[ D.A.]}, \bibinfo{author}{Clausen\xfnm[ J.R.]},
  \bibinfo{author}{Aidun\xfnm[ C.K.]}.
\newblock \bibinfo{title}{Coupling the lattice-boltzmann and spectrin-link
  methods for the direct numerical simulation of cellular blood flow}.
\newblock \bibinfo{journal}{International Journal for Numerical Methods in
  Fluids}
  \bibinfo{year}{2011};\bibinfo{volume}{68}(\bibinfo{number}{6}):\bibinfo{pages}{767--781}.
\newblock \DOIprefix\doi{10.1002/fld.2534}.
\bibitem[{Shan and Chen(1993)}]{Shan1993LatticeComponents}
\bibinfo{author}{Shan\xfnm[ X.]}, \bibinfo{author}{Chen\xfnm[ H.]}.
\newblock \bibinfo{title}{Lattice boltzmann model for simulating flows with
  multiple phases and components}.
\newblock \bibinfo{journal}{Physical Review E}
  \bibinfo{year}{1993};\bibinfo{volume}{47}(\bibinfo{number}{3}):\bibinfo{pages}{1815--1819}.
\newblock \DOIprefix\doi{10.1103/PhysRevE.47.1815}.
\bibitem[{Feng et~al.(2007)Feng, Han and Owen}]{Feng_LBM_ComputationalIssues}
\bibinfo{author}{Feng\xfnm[ Y.T.]}, \bibinfo{author}{Han\xfnm[ K.]},
  \bibinfo{author}{Owen\xfnm[ D.R.J.]}.
\newblock \bibinfo{title}{Coupled lattice boltzmann method and discrete element
  modelling of particle transport in turbulent fluid flows: Computational
  issues}.
\newblock \bibinfo{journal}{International Journal for Numerical Methods in
  Engineering}
  \bibinfo{year}{2007};\bibinfo{volume}{72}(\bibinfo{number}{9}):\bibinfo{pages}{1111--1134}.
\newblock \DOIprefix\doi{10.1002/nme.2114}.
\bibitem[{Kr{\"{u}}ger et~al.(2017)Kr{\"{u}}ger, Kusumaatmaja, Kuzmin, Shardt,
  Silva and Viggen}]{Kruger2017TheMethod}
\bibinfo{author}{Kr{\"{u}}ger\xfnm[ T.]}, \bibinfo{author}{Kusumaatmaja\xfnm[
  H.]}, \bibinfo{author}{Kuzmin\xfnm[ A.]}, \bibinfo{author}{Shardt\xfnm[ O.]},
  \bibinfo{author}{Silva\xfnm[ G.]}, \bibinfo{author}{Viggen\xfnm[ E.M.]}.
\newblock \bibinfo{title}{The Lattice Boltzmann Method}.
\newblock \bibinfo{year}{2017}.
\newblock \DOIprefix\doi{10.1007/978-3-319-44649-3}.
\bibitem[{Kr{\"{u}}ger(2012)}]{KrugerThesis}
\bibinfo{author}{Kr{\"{u}}ger\xfnm[ T.]}.
\newblock \bibinfo{title}{Computer Simulation Study of Collective Phenomena in
  Dense Suspensions of Red Blood Cells under Shear}.
\newblock \bibinfo{publisher}{Vieweg+Teubner Verlag}; \bibinfo{year}{2012}.
\newblock \DOIprefix\doi{10.1007/978-3-8348-2376-2}.
\bibitem[{Ota et~al.(2012)Ota, Suzuki and Inamuro}]{Ota2012LiftSimulations}
\bibinfo{author}{Ota\xfnm[ K.]}, \bibinfo{author}{Suzuki\xfnm[ K.]},
  \bibinfo{author}{Inamuro\xfnm[ T.]}.
\newblock \bibinfo{title}{Lift generation by a two-dimensional symmetric
  flapping wing: Immersed boundary-lattice boltzmann simulations}.
\newblock \bibinfo{journal}{Fluid Dynamics Research}
  \bibinfo{year}{2012};\bibinfo{volume}{44}(\bibinfo{number}{4}).
\newblock \DOIprefix\doi{10.1088/0169-5983/44/4/045504}.
\bibitem[{Mountrakis et~al.(2017)Mountrakis, Lorenz and
  Hoekstra}]{Mountrakis_2017_IBM}
\bibinfo{author}{Mountrakis\xfnm[ L.]}, \bibinfo{author}{Lorenz\xfnm[ E.]},
  \bibinfo{author}{Hoekstra\xfnm[ A.G.]}.
\newblock \bibinfo{title}{Revisiting the use of the immersed-boundary
  lattice-boltzmann method for simulations of suspended particles}.
\newblock \bibinfo{journal}{Phys Rev E}
  \bibinfo{year}{2017};\bibinfo{volume}{96}:\bibinfo{pages}{013302}.
\newblock \DOIprefix\doi{10.1103/PhysRevE.96.013302}.
\bibitem[{Peskin(1972)}]{Peskin1972FlowMethod}
\bibinfo{author}{Peskin\xfnm[ C.S.]}.
\newblock \bibinfo{title}{Flow patterns around heart valves: A numerical
  method}.
\newblock \bibinfo{journal}{Journal of Computational Physics}
  \bibinfo{year}{1972};\bibinfo{volume}{10}(\bibinfo{number}{2}):\bibinfo{pages}{252--271}.
\newblock \DOIprefix\doi{10.1016/0021-9991(72)90065-4}.
\bibitem[{Mountrakis et~al.(2015)Mountrakis, Lorenz, Malaspinas, Alowayyed,
  Chopard and Hoekstra}]{Mountrakis2015ParallelFramework}
\bibinfo{author}{Mountrakis\xfnm[ L.]}, \bibinfo{author}{Lorenz\xfnm[ E.]},
  \bibinfo{author}{Malaspinas\xfnm[ O.]}, \bibinfo{author}{Alowayyed\xfnm[
  S.]}, \bibinfo{author}{Chopard\xfnm[ B.]}, \bibinfo{author}{Hoekstra\xfnm[
  A.G.]}.
\newblock \bibinfo{title}{Parallel performance of an ib-lbm suspension
  simulation framework}.
\newblock \bibinfo{journal}{Journal of Computational Science}
  \bibinfo{year}{2015};\bibinfo{volume}{9}:\bibinfo{pages}{45--50}.
\newblock \DOIprefix\doi{10.1016/j.jocs.2015.04.006}.
\bibitem[{Zavodszky et~al.(2017)Zavodszky, van Rooij, Azizi, Alowayyed and
  Hoekstra}]{Zavodszky2017Hemocell:Library}
\bibinfo{author}{Zavodszky\xfnm[ G.]}, \bibinfo{author}{van Rooij\xfnm[ B.]},
  \bibinfo{author}{Azizi\xfnm[ V.]}, \bibinfo{author}{Alowayyed\xfnm[ S.]},
  \bibinfo{author}{Hoekstra\xfnm[ A.]}.
\newblock \bibinfo{title}{Hemocell: a high-performance microscopic cellular
  library}.
\newblock \bibinfo{journal}{Procedia Computer Science}
  \bibinfo{year}{2017};\bibinfo{volume}{108}:\bibinfo{pages}{159--165}.
\newblock \DOIprefix\doi{10.1016/j.procs.2017.05.084}.
\bibitem[{Tan et~al.(2018)Tan, Sinno and Diamond}]{Tan_2018_PALABOS_LAMMPS}
\bibinfo{author}{Tan\xfnm[ J.]}, \bibinfo{author}{Sinno\xfnm[ T.R.]},
  \bibinfo{author}{Diamond\xfnm[ S.L.]}.
\newblock \bibinfo{title}{A parallel fluid–solid coupling model using lammps
  and palabos based on the immersed boundary method}.
\newblock \bibinfo{journal}{Journal of Computational Science}
  \bibinfo{year}{2018};\bibinfo{volume}{25}:\bibinfo{pages}{89 -- 100}.
\newblock \DOIprefix\doi{10.1016/j.jocs.2018.02.006}.
\bibitem[{Sha(2014)}]{ShapeOpWebSite}
\bibinfo{title}{Shapeop}.
\newblock \bibinfo{howpublished}{\url{https://www.shapeop.org/}};
  \bibinfo{year}{2014}.
\bibitem[{{Bény} et~al.(2019){Bény}, {Kotsalos} and {Latt}}]{Beny_2019}
\bibinfo{author}{{Bény}\xfnm[ J.]}, \bibinfo{author}{{Kotsalos}\xfnm[ C.]},
  \bibinfo{author}{{Latt}\xfnm[ J.]}.
\newblock \bibinfo{title}{Toward full gpu implementation of fluid-structure
  interaction}.
\newblock In: \bibinfo{booktitle}{2019 18th International Symposium on Parallel
  and Distributed Computing (ISPDC)}. \bibinfo{year}{2019}, p.
  \bibinfo{pages}{16--22}.
\newblock \DOIprefix\doi{10.1109/ISPDC.2019.000-2}.
\bibitem[{Chopard et~al.(2017)Chopard, de~Sousa, Lätt, Mountrakis, Dubois,
  Yourassowsky et~al.}]{Chopard_PLTs_2017}
\bibinfo{author}{Chopard\xfnm[ B.]}, \bibinfo{author}{de~Sousa\xfnm[ D.R.]},
  \bibinfo{author}{Lätt\xfnm[ J.]}, \bibinfo{author}{Mountrakis\xfnm[ L.]},
  \bibinfo{author}{Dubois\xfnm[ F.]}, \bibinfo{author}{Yourassowsky\xfnm[ C.]},
  et~al.
\newblock \bibinfo{title}{A physical description of the adhesion and
  aggregation of platelets}.
\newblock \bibinfo{journal}{Royal Society Open Science}
  \bibinfo{year}{2017};\bibinfo{volume}{4}(\bibinfo{number}{4}):\bibinfo{pages}{170219}.
\newblock \DOIprefix\doi{10.1098/rsos.170219}.
\bibitem[{Blumers et~al.(2017)Blumers, Tang, Li, Li and
  Karniadakis}]{BLUMERS2017171}
\bibinfo{author}{Blumers\xfnm[ A.L.]}, \bibinfo{author}{Tang\xfnm[ Y.H.]},
  \bibinfo{author}{Li\xfnm[ Z.]}, \bibinfo{author}{Li\xfnm[ X.]},
  \bibinfo{author}{Karniadakis\xfnm[ G.E.]}.
\newblock \bibinfo{title}{Gpu-accelerated red blood cells simulations with
  transport dissipative particle dynamics}.
\newblock \bibinfo{journal}{Computer Physics Communications}
  \bibinfo{year}{2017};\bibinfo{volume}{217}:\bibinfo{pages}{171 -- 179}.
\newblock \DOIprefix\doi{10.1016/j.cpc.2017.03.016}.
\bibitem[{Clausen et~al.(2010)Clausen, Reasor and Aidun}]{Clausen_2010}
\bibinfo{author}{Clausen\xfnm[ J.R.]}, \bibinfo{author}{Reasor\xfnm[ D.A.]},
  \bibinfo{author}{Aidun\xfnm[ C.K.]}.
\newblock \bibinfo{title}{Parallel performance of a lattice-boltzmann/finite
  element cellular blood flow solver on the ibm blue gene/p architecture}.
\newblock \bibinfo{journal}{Computer Physics Communications}
  \bibinfo{year}{2010};\bibinfo{volume}{181}(\bibinfo{number}{6}):\bibinfo{pages}{1013
  -- 1020}.
\newblock \DOIprefix\doi{10.1016/j.cpc.2010.02.005}.
\bibitem[{Zydney and Colton(1988)}]{Zydney_Colton}
\bibinfo{author}{Zydney\xfnm[ A.]}, \bibinfo{author}{Colton\xfnm[ C.]}.
\newblock \bibinfo{title}{Augmented solute transport in the shear flow of a
  concentrated suspension.}
\newblock \bibinfo{journal}{International Journal of Multiphase Flow}
  \bibinfo{year}{1988};\bibinfo{volume}{10}(\bibinfo{number}{1}):\bibinfo{pages}{77--96}.
\bibitem[{Affeld et~al.(2013)Affeld, Goubergrits, Watanabe and
  Kertzscher}]{Affeld_2013}
\bibinfo{author}{Affeld\xfnm[ K.]}, \bibinfo{author}{Goubergrits\xfnm[ L.]},
  \bibinfo{author}{Watanabe\xfnm[ N.]}, \bibinfo{author}{Kertzscher\xfnm[ U.]}.
\newblock \bibinfo{title}{Numerical and experimental evaluation of platelet
  deposition to collagen coated surface at low shear rates}.
\newblock \bibinfo{journal}{Journal of Biomechanics}
  \bibinfo{year}{2013};\bibinfo{volume}{46}(\bibinfo{number}{2}):\bibinfo{pages}{430
  -- 436}.
\newblock \DOIprefix\doi{10.1016/j.jbiomech.2012.10.030};
  \bibinfo{note}{special Issue: Biofluid Mechanics}.
\bibitem[{Kumar and Graham(2011)}]{Kumar_Graham_2011}
\bibinfo{author}{Kumar\xfnm[ A.]}, \bibinfo{author}{Graham\xfnm[ M.D.]}.
\newblock \bibinfo{title}{Segregation by membrane rigidity in flowing binary
  suspensions of elastic capsules}.
\newblock \bibinfo{journal}{Phys Rev E}
  \bibinfo{year}{2011};\bibinfo{volume}{84}:\bibinfo{pages}{066316}.
\newblock \DOIprefix\doi{10.1103/PhysRevE.84.066316}.
\bibitem[{Kumar and Graham(2012)}]{Kumar_Graham_2012}
\bibinfo{author}{Kumar\xfnm[ A.]}, \bibinfo{author}{Graham\xfnm[ M.D.]}.
\newblock \bibinfo{title}{Margination and segregation in confined flows of
  blood and other multicomponent suspensions}.
\newblock \bibinfo{journal}{Soft Matter}
  \bibinfo{year}{2012};\bibinfo{volume}{8}:\bibinfo{pages}{10536--10548}.
\newblock \DOIprefix\doi{10.1039/C2SM25943E}.
\bibitem[{Mountrakis et~al.(2013)Mountrakis, Lorenz and
  Hoekstra}]{Mountrakis_whereDoPLTsGo}
\bibinfo{author}{Mountrakis\xfnm[ L.]}, \bibinfo{author}{Lorenz\xfnm[ E.]},
  \bibinfo{author}{Hoekstra\xfnm[ A.G.]}.
\newblock \bibinfo{title}{Where do the platelets go? a simulation study of
  fully resolved blood flow through aneurysmal vessels}.
\newblock \bibinfo{journal}{Interface Focus}
  \bibinfo{year}{2013};\bibinfo{volume}{3}(\bibinfo{number}{2}):\bibinfo{pages}{20120089}.
\newblock \DOIprefix\doi{10.1098/rsfs.2012.0089}.
\bibitem[{{Herschlag} et~al.(2019){Herschlag}, {Gounley}, {Roychowdhury},
  {Draeger} and {Randles}}]{Randles_2019}
\bibinfo{author}{{Herschlag}\xfnm[ G.]}, \bibinfo{author}{{Gounley}\xfnm[ J.]},
  \bibinfo{author}{{Roychowdhury}\xfnm[ S.]}, \bibinfo{author}{{Draeger}\xfnm[
  E.]}, \bibinfo{author}{{Randles}\xfnm[ A.]}.
\newblock \bibinfo{title}{Multi-physics simulations of particle tracking in
  arterial geometries with a scalable moving window algorithm}.
\newblock In: \bibinfo{booktitle}{IEEE Cluster 2019}.
  \bibinfo{address}{Albuquerque, New Mexico USA}: \bibinfo{publisher}{IEEE};
  \bibinfo{year}{2019},.

\end{thebibliography}


\begin{thebibliography}{14}
\providecommand{\natexlab}[1]{#1}
\providecommand{\url}[1]{\texttt{#1}}
\providecommand{\href}[2]{#2}
\providecommand{\path}[1]{#1}
\providecommand{\eprint}[1]{\href{http://arxiv.org/abs/#1}{\path{#1}}}
\providecommand{\DOIprefix}{doi:}
\providecommand{\ArXivprefix}{arXiv:}
\providecommand{\URLprefix}{URL: }
\providecommand{\Pubmedprefix}{pmid:}
\providecommand{\doi}[1]{\href{http://dx.doi.org/#1}{\path{#1}}}
\providecommand{\Pubmed}[1]{\href{pmid:#1}{\path{#1}}}
\providecommand{\BIBand}{and}
\providecommand{\bibinfo}[2]{#2}
\ifx\xfnm\undefined \def\xfnm[#1]{\unskip,\space#1}\fi
\bibitem[{Kr{\"{u}}ger et~al.(2017)Kr{\"{u}}ger, Kusumaatmaja, Kuzmin, Shardt,
  Silva and Viggen}]{Kruger2017TheMethod}
\bibinfo{author}{Kr{\"{u}}ger\xfnm[ T.]}, \bibinfo{author}{Kusumaatmaja\xfnm[
  H.]}, \bibinfo{author}{Kuzmin\xfnm[ A.]}, \bibinfo{author}{Shardt\xfnm[ O.]},
  \bibinfo{author}{Silva\xfnm[ G.]}, \bibinfo{author}{Viggen\xfnm[ E.M.]}.
\newblock \bibinfo{title}{The Lattice Boltzmann Method}.
\newblock \bibinfo{year}{2017}.
\newblock \DOIprefix\doi{10.1007/978-3-319-44649-3}.
\bibitem[{Latt et~al.(2019)Latt, Malaspinas, Kontaxakis, Parmigiani, Lagrava,
  Brogi et~al.}]{PalabosArticle}
\bibinfo{author}{Latt\xfnm[ J.]}, \bibinfo{author}{Malaspinas\xfnm[ O.]},
  \bibinfo{author}{Kontaxakis\xfnm[ D.]}, \bibinfo{author}{Parmigiani\xfnm[
  A.]}, \bibinfo{author}{Lagrava\xfnm[ D.]}, \bibinfo{author}{Brogi\xfnm[ F.]},
  et~al.
\newblock \bibinfo{title}{Palabos: Parallel lattice boltzmann solver}
  \bibinfo{year}{2019};\DOIprefix\doi{10.13140/RG.2.2.20836.94086}.
\bibitem[{Pal(2019)}]{PalabosWebsite}
\bibinfo{title}{Palabos}.
\newblock \bibinfo{howpublished}{\url{https://palabos.unige.ch/}};
  \bibinfo{year}{2019}.
\bibitem[{Kotsalos et~al.(2019)Kotsalos, Latt and
  Chopard}]{Kotsalos_JCP_2019_npFEM}
\bibinfo{author}{Kotsalos\xfnm[ C.]}, \bibinfo{author}{Latt\xfnm[ J.]},
  \bibinfo{author}{Chopard\xfnm[ B.]}.
\newblock \bibinfo{title}{Bridging the computational gap between mesoscopic and
  continuum modeling of red blood cells for fully resolved blood flow}.
\newblock \bibinfo{journal}{Journal of Computational Physics}
  \bibinfo{year}{2019};\bibinfo{volume}{398}.
\newblock \DOIprefix\doi{10.1016/j.jcp.2019.108905}.
\bibitem[{Sha(2014)}]{ShapeOpWebSite}
\bibinfo{title}{Shapeop}.
\newblock \bibinfo{howpublished}{\url{https://www.shapeop.org/}};
  \bibinfo{year}{2014}.
\bibitem[{M{\"{u}}ller et~al.(2007)M{\"{u}}ller, Heidelberger, Hennix and
  Ratcliff}]{Muller2007PositionDynamics}
\bibinfo{author}{M{\"{u}}ller\xfnm[ M.]}, \bibinfo{author}{Heidelberger\xfnm[
  B.]}, \bibinfo{author}{Hennix\xfnm[ M.]}, \bibinfo{author}{Ratcliff\xfnm[
  J.]}.
\newblock \bibinfo{title}{Position based dynamics}.
\newblock \bibinfo{journal}{Journal of Visual Communication and Image
  Representation}
  \bibinfo{year}{2007};\bibinfo{volume}{18}(\bibinfo{number}{2}):\bibinfo{pages}{109--118}.
\newblock \DOIprefix\doi{10.1016/j.jvcir.2007.01.005}.
\bibitem[{Feng et~al.(2007)Feng, Han and Owen}]{Feng_LBM_ComputationalIssues}
\bibinfo{author}{Feng\xfnm[ Y.T.]}, \bibinfo{author}{Han\xfnm[ K.]},
  \bibinfo{author}{Owen\xfnm[ D.R.J.]}.
\newblock \bibinfo{title}{Coupled lattice boltzmann method and discrete element
  modelling of particle transport in turbulent fluid flows: Computational
  issues}.
\newblock \bibinfo{journal}{International Journal for Numerical Methods in
  Engineering}
  \bibinfo{year}{2007};\bibinfo{volume}{72}(\bibinfo{number}{9}):\bibinfo{pages}{1111--1134}.
\newblock \DOIprefix\doi{10.1002/nme.2114}.
\bibitem[{Bhatnagar et~al.(1954)Bhatnagar, Gross and
  Krook}]{Bhatnagar1954ASystems}
\bibinfo{author}{Bhatnagar\xfnm[ P.L.]}, \bibinfo{author}{Gross\xfnm[ E.P.]},
  \bibinfo{author}{Krook\xfnm[ M.]}.
\newblock \bibinfo{title}{A model for collision processes in gases. small
  amplitude processes in charged and neutral one-component systems}.
\newblock \bibinfo{journal}{Physical Review}
  \bibinfo{year}{1954};\bibinfo{volume}{94}(\bibinfo{number}{3}):\bibinfo{pages}{511--525}.
\newblock \DOIprefix\doi{10.1103/PhysRev.94.511}.
\bibitem[{Shan and Chen(1993)}]{Shan1993LatticeComponents}
\bibinfo{author}{Shan\xfnm[ X.]}, \bibinfo{author}{Chen\xfnm[ H.]}.
\newblock \bibinfo{title}{Lattice boltzmann model for simulating flows with
  multiple phases and components}.
\newblock \bibinfo{journal}{Physical Review E}
  \bibinfo{year}{1993};\bibinfo{volume}{47}(\bibinfo{number}{3}):\bibinfo{pages}{1815--1819}.
\newblock \DOIprefix\doi{10.1103/PhysRevE.47.1815}.
\bibitem[{Kr{\"{u}}ger(2012)}]{KrugerThesis}
\bibinfo{author}{Kr{\"{u}}ger\xfnm[ T.]}.
\newblock \bibinfo{title}{Computer Simulation Study of Collective Phenomena in
  Dense Suspensions of Red Blood Cells under Shear}.
\newblock \bibinfo{publisher}{Vieweg+Teubner Verlag}; \bibinfo{year}{2012}.
\newblock \DOIprefix\doi{10.1007/978-3-8348-2376-2}.
\bibitem[{Mittal and Iaccarino(2005)}]{Mittal2005ImmersedMethods}
\bibinfo{author}{Mittal\xfnm[ R.]}, \bibinfo{author}{Iaccarino\xfnm[ G.]}.
\newblock \bibinfo{title}{Immersed boundary methods}.
\newblock \bibinfo{journal}{Annual Review of Fluid Mechanics}
  \bibinfo{year}{2005};\bibinfo{volume}{37}(\bibinfo{number}{1}):\bibinfo{pages}{239--261}.
\newblock \DOIprefix\doi{10.1146/annurev.fluid.37.061903.175743}.
\bibitem[{Wang et~al.(2008)Wang, Fan and Luo}]{Wang2008CombinedParticles}
\bibinfo{author}{Wang\xfnm[ Z.]}, \bibinfo{author}{Fan\xfnm[ J.]},
  \bibinfo{author}{Luo\xfnm[ K.]}.
\newblock \bibinfo{title}{Combined multi-direct forcing and immersed boundary
  method for simulating flows with moving particles}.
\newblock \bibinfo{journal}{International Journal of Multiphase Flow}
  \bibinfo{year}{2008};\bibinfo{volume}{34}(\bibinfo{number}{3}):\bibinfo{pages}{283--302}.
\newblock \DOIprefix\doi{10.1016/j.ijmultiphaseflow.2007.10.004}.
\bibitem[{Ota et~al.(2012)Ota, Suzuki and Inamuro}]{Ota2012LiftSimulations}
\bibinfo{author}{Ota\xfnm[ K.]}, \bibinfo{author}{Suzuki\xfnm[ K.]},
  \bibinfo{author}{Inamuro\xfnm[ T.]}.
\newblock \bibinfo{title}{Lift generation by a two-dimensional symmetric
  flapping wing: Immersed boundary-lattice boltzmann simulations}.
\newblock \bibinfo{journal}{Fluid Dynamics Research}
  \bibinfo{year}{2012};\bibinfo{volume}{44}(\bibinfo{number}{4}).
\newblock \DOIprefix\doi{10.1088/0169-5983/44/4/045504}.
\bibitem[{Mountrakis et~al.(2017)Mountrakis, Lorenz and
  Hoekstra}]{Mountrakis_2017_IBM}
\bibinfo{author}{Mountrakis\xfnm[ L.]}, \bibinfo{author}{Lorenz\xfnm[ E.]},
  \bibinfo{author}{Hoekstra\xfnm[ A.G.]}.
\newblock \bibinfo{title}{Revisiting the use of the immersed-boundary
  lattice-boltzmann method for simulations of suspended particles}.
\newblock \bibinfo{journal}{Phys Rev E}
  \bibinfo{year}{2017};\bibinfo{volume}{96}:\bibinfo{pages}{013302}.
\newblock \DOIprefix\doi{10.1103/PhysRevE.96.013302}.

\end{thebibliography}

\end{document}


\begin{frontmatter}

\title{Digital Blood in Massively Parallel CPU/GPU Systems for the Study of Platelet Transport \\
\textbf{Supplementary Material}}

\author[UniGe]{Christos Kotsalos \corref{correspondingauthor}}
\ead{christos.kotsalos@unige.ch}

\author[UniGe]{Jonas Latt}

\author[UniGe]{Joel Beny}

\author[UniGe]{Bastien Chopard}

\address[UniGe]{Computer Science Department, University of Geneva, 7 route de Drize, CH-1227 Carouge, Switzerland}

\cortext[correspondingauthor]{Corresponding Author}

\begin{abstract}
We propose a highly versatile computational framework for the simulation of cellular blood flow focusing on extreme performance without compromising accuracy or complexity. The tool couples the lattice Boltzmann solver Palabos for the simulation of the blood plasma, a novel finite element method (FEM) solver for the resolution of the deformable blood cells, and an immersed boundary method for the coupling of the two phases. The design of the tool supports hybrid CPU-GPU executions (fluid, fluid-solid interaction on CPUs, the FEM solver on GPUs), and is non-intrusive, as each of the three components can be replaced in a modular way. The FEM-based kernel for solid dynamics outperforms other FEM solvers and its performance is comparable to the state-of-the-art mass-spring systems. We perform an exhaustive performance analysis on Piz Daint at the Swiss National Supercomputing Centre and provide case studies focused on platelet transport. The tests show that this versatile framework combines unprecedented accuracy with massive performance, rendering it suitable for the upcoming exascale architectures.
\end{abstract}

\begin{keyword}
npFEM \sep Palabos \sep GPUs \sep Cellular Blood Flow \sep Platelet Transport
\end{keyword}

\end{frontmatter}

\section{Parameters used for the simulations}
\begin{table}[H]
\color{black}
\centering
\begin{tabular}{lc}
\hline
\textbf{Parameter} & \textbf{Value} \\
\hline
$\Delta x$ & $0.5~\mu m$ \\
relaxation time $\tau$ & $2$ \\
$\Delta t$ & $\nu = C_s^2 (\tau - 1/2) \Delta t$ \\
Skalak B,C,D & $5, 5000, 35~pN/\mu m$ \\
$k_{bending}$ & $1~pN\mu m$\\
$k_{volume}$ & $250$ \\
Rayleigh $\alpha_D, \beta_D$ & $0, 0.01$\\
$\kappa_{\text{damping}}$ & $\left \{ 0.5,0.7,0.8,0.9 \right \}$ \\
IBM cycles & $1$ \\
RBC mesh resolution (surface vertices) & $258$ \\
Platelet mesh resolution (surface vertices) & $66$ \\
Fluid density, kinematic viscosity & $1025~fg/\mu m^3, 1~\mu m^2/\mu s$ \\
\end{tabular}
\label{tab:Params}
\end{table}

\section{Methods}
The computational framework used in this study simulates the particulate nature of blood, i.e. the blood cells that are submerged in the blood plasma. More in details, we suggest a high-performance computational framework for fully resolved 3D blood flow simulations. The tool is based on three computational modules, namely the fluid solver, the solid solver and the fluid-solid interaction (FSI).

The fluid solver is based on the lattice Boltzmann method (LBM) \cite{Kruger2017TheMethod} and the module uses Palabos \cite{PalabosArticle}, \cite{PalabosWebsite} as for the implementation of the LBM. Palabos stands for Parallel Lattice Boltzmann Solver and it is an open source software maintained by the Scientific and Parallel Computing Group (SPC) at the University of Geneva. The FSI is done via an immersed boundary method known as multidirect forcing scheme and its implementation is realised in Palabos. Lastly, the solid solver is based on the nodal projective finite elements method (npFEM) \cite{Kotsalos_JCP_2019_npFEM} and the module uses the npFEM library developed and maintained by the Palabos team, based on the open-source ShapeOp library \cite{ShapeOpWebSite}. In a future release of Palabos, we are planning on integrating npFEM on the core library.

In this section, a basic and brief outline of the used methods is given for the sake of completeness of the article. For a thorough presentation of the numerical methods, the interested readers should consult Kotsalos \emph{et al.} \cite{Kotsalos_JCP_2019_npFEM}.

\subsection{Nodal Projective FEM (npFEM)}
The deformable blood cells are simulated by the nodal projective finite elements method. The npFEM is a mass-lumped linear FE solver that resolves both the trajectories and deformations of the blood cells with high accuracy. The solver has the capability of capturing the rich and non-linear viscoelastic behaviour of red blood cells as shown and validated in \cite{Kotsalos_JCP_2019_npFEM}. The rest of the blood cells, platelets for the current study, are simulated as nearly-rigid bodies by increasing the stiffness of the material in the solid solver.  

Let us assume a surface mesh consisting of $n$ vertices with positions $\bm{x} \in \mathbb{R}^{n \times 3}$ and velocities $\bm{v} \in \mathbb{R}^{n \times 3}$. The evolution of the body (trajectory \& deformed shape) in time follows Newton's laws of motion. At time $t_n$, the system is described by $\{ \bm{x}_n, \bm{v}_n \}$. The external forces are defined as $\bm{F}_{ext} \in \mathbb{R}^{n \times 3}$ and are due to the fluid-solid interaction, while the internal forces are $\bm{F}_{int} \in \mathbb{R}^{n \times 3}$ and are due to the material properties. The internal forces are given by $\bm{F}_{int}(\bm{x}) = - \sum_{i} \nabla E_i(\bm{x})$, where $E_i(\bm{x})$ is a scalar discrete elemental potential energy (the summation of all the elemental potential energies results in the total elastic potential energy of the body). The potential energy describes how the body reacts to the imposition of external forces. The \emph{implicit Euler time integration} leads to the following advancement rules:
\begin{flalign}
    \bm{v}_{n+1} = \frac{\bm{x}_{n+1} - \bm{x}_{n}}{h}, \label{eq:vel}   \\
    \bm{F}_{int}(\bm{x}_{n+1}) + \bm{F}_{ext}(\bm{x}_{n}) - \mathbf{C} \bm{v}_{n+1} = \mathbf{M} \frac{\bm{v}_{n+1} - \bm{v}_{n}}{h}, \label{eq:Newton2nd} &&
\end{flalign}
where $\mathbf{M} \in \mathbb{R}^{n \times n}$ is the mass matrix, $h$ is the time step, subscripts $n$ and $n+1$ refer to time $t$ and $t+h$, respectively and $\mathbf{C} \in \mathbb{R}^{n \times n}$ is the damping matrix acting like a proxy for viscoelasticity. Except for the Rayleigh damping, we augment the viscoelastic behaviour through another proxy called the position-based dynamics damping \cite{Muller2007PositionDynamics} (more details in \cite{Kotsalos_JCP_2019_npFEM}). The mass matrix is built by lumping the total mass of the body (including the interior fluid) on the mesh vertices, resulting in a diagonal structure. We should mention as well that the implicit nature of the npFEM solver renders it capable of resolving very high deformations with unconditional stability for arbitrary time steps. Other integration schemes, such as the explicit Runge-Kutta, have been tested for even higher accuracy, but since we are interested in investigating the collective behaviour of multiple blood cells, this enhanced accuracy in modelling the trajectory of a single blood cell did not affect the overall behaviour. Equations \eqref{eq:vel} \& \eqref{eq:Newton2nd} can be combined and converted into an optimisation problem as
\begin{flalign} \label{eq:genericOptimizationNewton}
    \underset{\bm{x}_{n+1}}{\min}~~g(\bm{x}_{n+1}) = \frac{1}{2h^2} \left \| \widetilde{\mathbf{M}}^{\frac{1}{2}} (\bm{x}_{n+1} - \bm{y}_n) \right \| ^{2}_{F} + \sum_{i} E_i(\bm{x}_{n+1}), &&
\end{flalign}
where $\widetilde{\mathbf{M}} = \mathbf{M} + h \mathbf{C}$, $\bm{y}_n = \bm{x}_{n} + h \widetilde{\mathbf{M}}^{-1} \mathbf{M} \bm{v}_n + h^2 \widetilde{\mathbf{M}}^{-1} \bm{F}_{ext}$ and $|| \cdot ||_F$ is the Frobenius norm. Indeed, setting the derivative of equation \eqref{eq:genericOptimizationNewton} to zero (thereby minimising the objective function), we recover Newton's second law of motion. The solution to the above optimisation problem gives the system's state at the next time step, i.e. $\{ \bm{x}_{n+1}, \bm{v}_{n+1} \}$ and thus resolves the trajectory and deformed state of the body at time $t+h$. The steps to minimise equation \eqref{eq:genericOptimizationNewton} are thoroughly described in Kotsalos \emph{et al.} \cite{Kotsalos_JCP_2019_npFEM}, but roughly a quasi-Newton approach is deployed that converges to $\bm{x}_{n+1}$ through a sequence of iterations ($k=1,...$) as
\begin{flalign}
    \bm{x}_{n+1}^{k+1} \leftarrow \bm{x}_{n+1}^{k} - \widetilde{\mathbf{H}} \nabla g(\bm{x}_{n+1}^{k}), &&
\end{flalign}
where $\widetilde{\mathbf{H}}$ is an approximation of the Hessian of $g(\bm{x}_{n+1})$. The Hessian approximation is built on top of the observation that most of the potential energies that describe the behaviour of a red blood cell are of quadratic form \cite{Kotsalos_JCP_2019_npFEM}.

The theory introduced above is not restricted to membranes only, but it can straightforwardly resolve fully 3D bodies, i.e. the ones discretized by tetrahedra. This section gave a broad overview of the topic and the main novelty of this approach is a cleaner way to describe elasticity through a variational point of view.

\subsection{Lattice Boltzmann Methods (LBM)}
The lattice Boltzmann method (LBM) is employed to solve indirectly the Navier-Stokes equations. Virtual particles, also known as populations, move on a regular grid/ lattice and collide at the lattice nodes. LBM is a second-order accurate solver for the weakly compressible Navier-Stokes equation, where the weak compressibility refers to errors that amplify as Mach number tends to one \citep{Kruger2017TheMethod}.

Let us consider a 3D incompressible fluid flow, with density $\rho$ and kinematic viscosity $\nu$. The fluid domain is discretised into a regular grid with spacing $\Delta x$ in all directions. The fluid is represented as a group of fictitious particles residing at lattice sites and move to the neighbouring nodes along a fixed set of discrete directions with given discrete velocities at discrete time steps $\Delta t$ \cite{Feng_LBM_ComputationalIssues}. For this study, we use a D3Q19 stencil and the fluid populations at each site are allowed to move to its 19 immediate neighbours with 19 different velocities $\left \{ \bm{c}_i  \right \}_{i=0}^{18}$. The degrees of freedom of the model are the populations, a group of 19 variables $\left \{ \bm{f}_i  \right \}_{i=0}^{18}$ on each cell, also know as fluid density distribution functions, each relating to the probable number of fluid particles moving along the $i$th direction. The evolution of the fluid through the discrete time is realised in two steps, i.e. the collision and streaming. The collision refers to the redistribution of the fluid populations at each lattice site and the streaming is their movement to neighbouring sites. By discretising the Boltzmann equation (which describes the evolution of the continuous distribution function $f$) in velocity space, physical space and time, we get the lattice Boltzmann equation (LBE)
\begin{flalign} \label{eq:LBE}
    f_i(\bm{x}, t) \gets f_i(\bm{x}, t) + \Omega_i(\bm{x}, t)~~~~~&(\text{Collision})\\
    f_i(\bm{x}+\bm{c}_i\Delta t, t+\Delta t) = f_i(\bm{x}, t)~~~~~&(\text{Streaming}),
\end{flalign}
which describes the collision and propagation of the mesoscopic particle packets. In all our simulations we use the linearised single-relaxation-time BGK \cite{Bhatnagar1954ASystems} formulation for $\Omega_i$. The macroscopic properties of the fluid such as the density $\rho$, the velocity $\bm{u}$ and the hydrodynamic stress tensor $\bm{\sigma}$ can directly be recovered at each lattice node and per time step from the moments of the distribution functions $f_i$. External forcing terms (like the FSI force $f_{imm}$) can be incorporated in the LBM through a modification of the collision term $\Omega$ using the Shan-Chen forcing scheme \cite{Shan1993LatticeComponents}.

Important quantities to consider in LBM are the lattice speed $C = \Delta x/ \Delta t$, the relaxation time $\tau$, the fluid speed of sound $C_s = C/\sqrt{3}$ and their coupling through the physical kinematic viscosity $\nu$ as
\begin{flalign} \label{eq:diffusiveScaling}
    \nu = C_s^2 (\tau - 1/2) \Delta t. &&
\end{flalign}
A particular point of interest is that in lattice Boltzmann simulations, if one fixes the spatial discretization and the relaxation time for a given fluid (known viscosity), then there is no freedom over the selection of the time discretization, as it is dictated by the diffusive scaling formula \eqref{eq:diffusiveScaling}. If $\Delta t$ is not defined through equation \eqref{eq:diffusiveScaling}, then physics are violated.

It can be proved that the updating rules (with BGK for $\Omega$) \eqref{eq:LBE} recover the incompressible Navier-Stokes to second order in both space and time. Since the LBM is obtained by the linearised expansion of the original kinetic theory-based LB equation, the resulting macroscopic variables converge to the Navier-Stokes equations with order $Ma^2$ \cite{Feng_LBM_ComputationalIssues}, where $Ma = u_{max} / C$ is the computational Mach number and $u_{max}$ is the maximum simulated velocity in the flow. Therefore for numerical accuracy and stability, it is required that $Ma \ll 1$.

For a complete overview of the lattice Boltzmann method, the reader should consult the work done by Kr{\"{u}}ger \emph{et al.} \cite{Kruger2017TheMethod}, \cite{KrugerThesis} \& Feng \emph{et al.} \cite{Feng_LBM_ComputationalIssues}.

\subsection{Immersed Boundary Method (IBM)}
The coupling between the blood cells (solid phase) and the blood plasma (fluid phase) is realised through the immersed boundary method (IBM). Essentially, the IBM imposes a no-slip boundary condition, so that each point of the surface and the ambient fluid move with the same velocity. The key idea is that the surface of the deformable object is viewed as a set of Lagrangian points, which do not have to conform with the fluid mesh/ lattice. The fluid feels the presence of the body only via a force field $f_{imm}$. The interaction between the off-lattice marker points (Lagrangian points) and the fluid is done via interpolation stencils, i.e. regularised Dirac delta functions. There are many variations of the IBM \cite{Mittal2005ImmersedMethods} and in this study we are using a modified version of the multi-direct forcing method introduced by Wang and colleagues \cite{Wang2008CombinedParticles}, in a form close to the one proposed by Ota et al. \cite{Ota2012LiftSimulations}.

The goal is to compute a forcing term $f_{imm}$ along the immersed boundaries and apply it as body force to the fluid (Shan-Chen forcing scheme). Let us denote with $\bm{x}$ and $\bm{X}_k$ the Eulerian and Lagrangian points respectively, with $\bm{U}_k$ the velocity of the Lagrangian point $k$ as computed by the solid solver (npFEM) and with $\bm{u}^*(\bm{x}, t)$ the fluid velocity as recovered by the distribution functions $f_i$. Using an interpolation stencil ($W$), the velocity on the Lagrangian point $\bm{X}_k$ is given by
\begin{flalign}
    \bm{u}^*(\bm{X}_k, t) = \sum_{\bm{x}} \bm{u}^*(\bm{x}, t) W(\bm{x} - \bm{X}_k) \Delta x^3, &&
\end{flalign}
where the interpolation stencil is the $\phi_4$ with support $4\Delta x$ \cite{Mountrakis_2017_IBM}.

The body force $\bm{f}_{imm}$ is determined by the following iterative procedure:
\begin{enumerate}
    \item[\textbf{Step 0.}] Initial estimation of the body force on the Lagrangian points by
        \begin{equation}
            \bm{f}_0(\bm{X}_k, t) = \frac{\bm{U}_k(t) - \bm{u}^*(\bm{X}_k, t)}{\Delta t} A_k,
        \end{equation}
        where $A_k$ is the corresponding surface area of the Lagrangian point $k$.
    \item[\textbf{Step 1.}] Distribute the force estimate from the Lagrangian point $\bm{X}_k$ to the neighbouring lattice sites at a range dictated by the interpolation kernel $\phi_4$. Correct the fluid velocity as $\bm{u}^* + \bm{f}_l \Delta t$ and re-interpolate the corrected velocity at the Lagrangian points.
    \item[\textbf{Step 2.}] Re-update the body force on Lagrangian point $\bm{X}_k$
        \begin{equation}
            \bm{f}_{l+1}(\bm{X}_k, t) = \bm{f}_l(\bm{X}_k, t) + \frac{\bm{U}_k(t) - \bm{u}_l(\bm{X}_k, t)}{\Delta t}  A_k,
        \end{equation}
        and repeat from \textbf{Step 1}.
\end{enumerate}
The above iterative procedure can be applied for as many cycles as the application demands. For simulations with multiple blood cells, even one cycle gives satisfactory results. The body force computed by the last cycle on the Eulerian grid is the $\bm{f}_{imm}$.

The immersed body force term $\bm{f}_{imm}$ could be used as the $\bm{F}_{ext}$ for the solid solver. However, we make use of the hydrodynamic stress tensor for higher accuracy. To compute the force on the Lagrangian point $k$, we project the stress tensor $\bm{\sigma}$ onto the surface normal $\bm{n}_k$.

We should highlight that in the classic immersed boundary method the same fluid covers the whole computational domain. However, this contradicts with the existence of interior fluid in the blood cells, of different viscosity and density, than the exterior blood plasma. Our approach is to completely disregard the interior fluid and view it as an artefact of the IBM. To do so, we compute the $\bm{F}_{ext}$ on the solid phase by summing only the points $\bm{x}$ that are on the outside region of the bodies. The effect of the interior fluid of the blood cells is implicitly taken into account from the npFEM solver by considering both its incompressibility and viscosity as additional terms to the viscoelasticity of the membrane. Thus, the bodies are simulated as entities that encapsulate all the different components of the blood cells \cite{Kotsalos_JCP_2019_npFEM}.

\subsection{Particle In Kernel (PIK) technique}
\begin{figure}[H]
    \centering
    \includegraphics[scale=0.7]{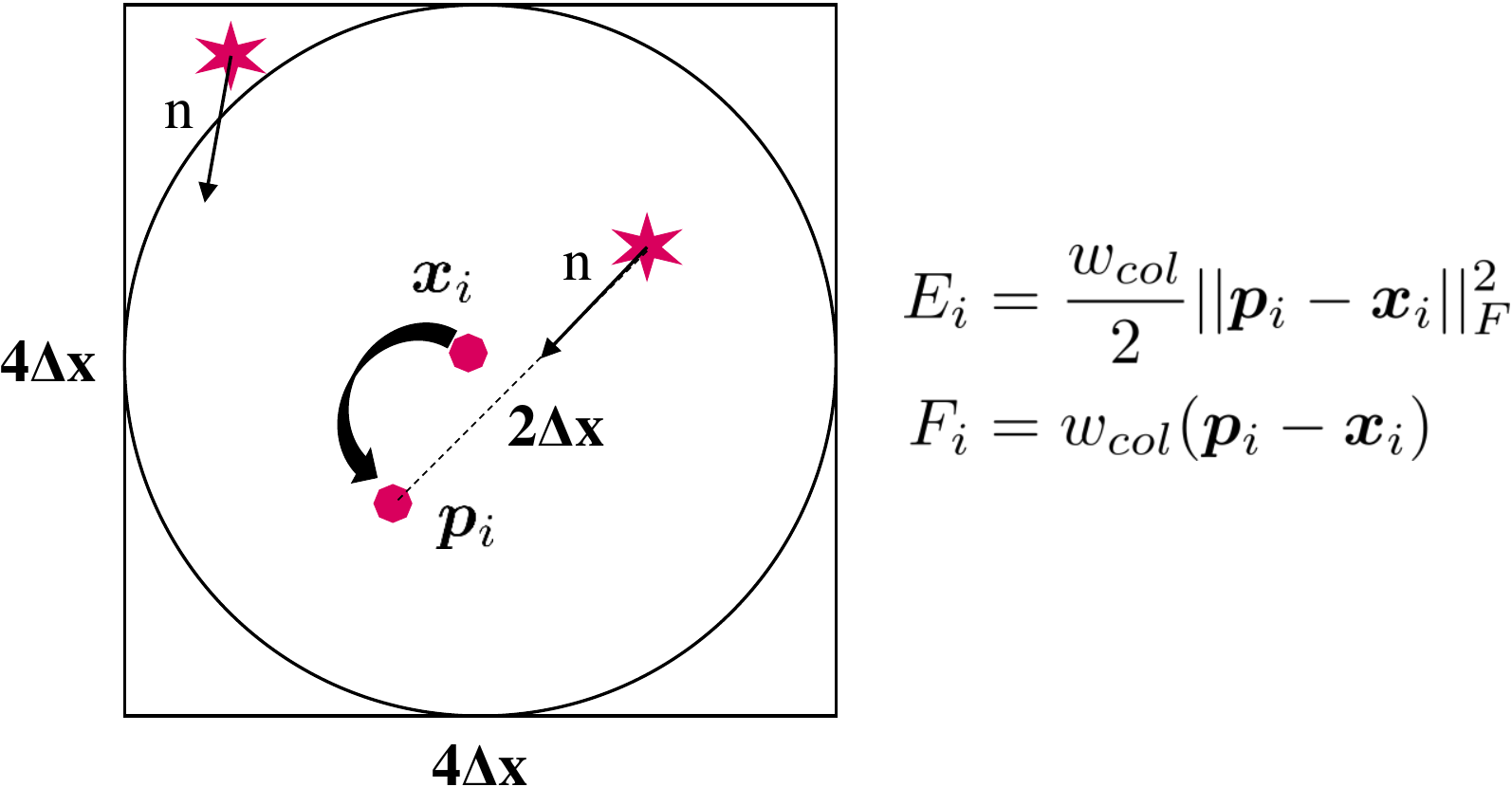}
    \caption{Particle In Kernel (PIK) technique and corresponding energy/force pair introduced in the npFEM solver.}
    \label{fig:PIK}
\end{figure}

The correct handling of colliding bodies, especially for flows at high hematocrit, is of paramount importance for accuracy and stability reasons. Bodies that approach each other very closely activate the collision framework which has to replace the hydrodynamic interactions because of under-resolution of the latter. Additionally, the IBM and the computation of the external forces on the immersed bodies make use of interpolation kernels that are drastically affected when they are ``contaminated'' by ``foreign'' particles. This means that the interpolated values consider regions that are not well-resolved, e.g. the interior fluid of the blood cells, and this could potentially lead to accuracy and stability problems. The Particle In Kernel (PIK) technique is essentially part of the collision framework and makes sure that the correct forces are used at all times. In Figure \ref{fig:PIK} there is a visual description of PIK for an investigated point $\bm{x}_i$ (dot) and 2 potential collision candidates (stars). If there is no ``foreign'' particle, i.e. particle belonging to another body, inside the kernel ($4\Delta x$ sphere in 3D) then the fluid resolution is sufficient to handle the flow field. Otherwise, we disregard the fluid force and the collision framework gives the force instead. The closest ``foreign'' particle is chosen to be the colliding neighbour and the desired location of $\bm{x}_i$ is noted as $\bm{p}_i$ (projection). Subsequently, a collision potential energy and a corresponding force is introduced in the solid solver. This approach guarantees to use only non-contaminated interpolation kernels and a clean collision framework. This technique affects only the computation of $\bm{F}_{ext}$ on the immersed boundaries and has no impact on the IBM steps.

\section{Piz Daint @ CSCS}
Piz Daint is the current flagship system of the Swiss National Supercomputing Centre (CSCS). This supercomputer is a hybrid Cray XC40/XC50 system. According to the list TOP500 (June 2019), Piz Daint is ranked $6^{th}$ worldwide, while it is the most powerful system in Europe. The specifications of the supercomputer are the following:

\begin{table}[H]
\centering
\begin{tabular}{|l|p{12cm}|}
\hline
Model & Cray XC40/XC50 \\\hline
XC50 Compute Nodes & Intel Xeon E5-2690 v3 @ 2.60GHz (12 cores, 64GB RAM) and NVIDIA Tesla P100 16GB - 5704 Nodes \\ \hline
XC40 Compute Nodes & Two Intel Xeon E5-2695 v4 @ 2.10GHz (2 x 18 cores, 64/128 GB RAM) - 1813 Nodes \\ \hline
Login Nodes & Intel Xeon CPU E5-2650 v3 @ 2.30GHz (10 cores, 256 GB RAM) \\ \hline
Interconnect Configuration & Aries routing and communications ASIC, and Dragonfly network topology \\ \hline
Scratch capacity & 8.8 PB \\ \hline
\end{tabular}
\end{table}

\begin{figure}[h]
    \centering
    \includegraphics[scale=1.0]{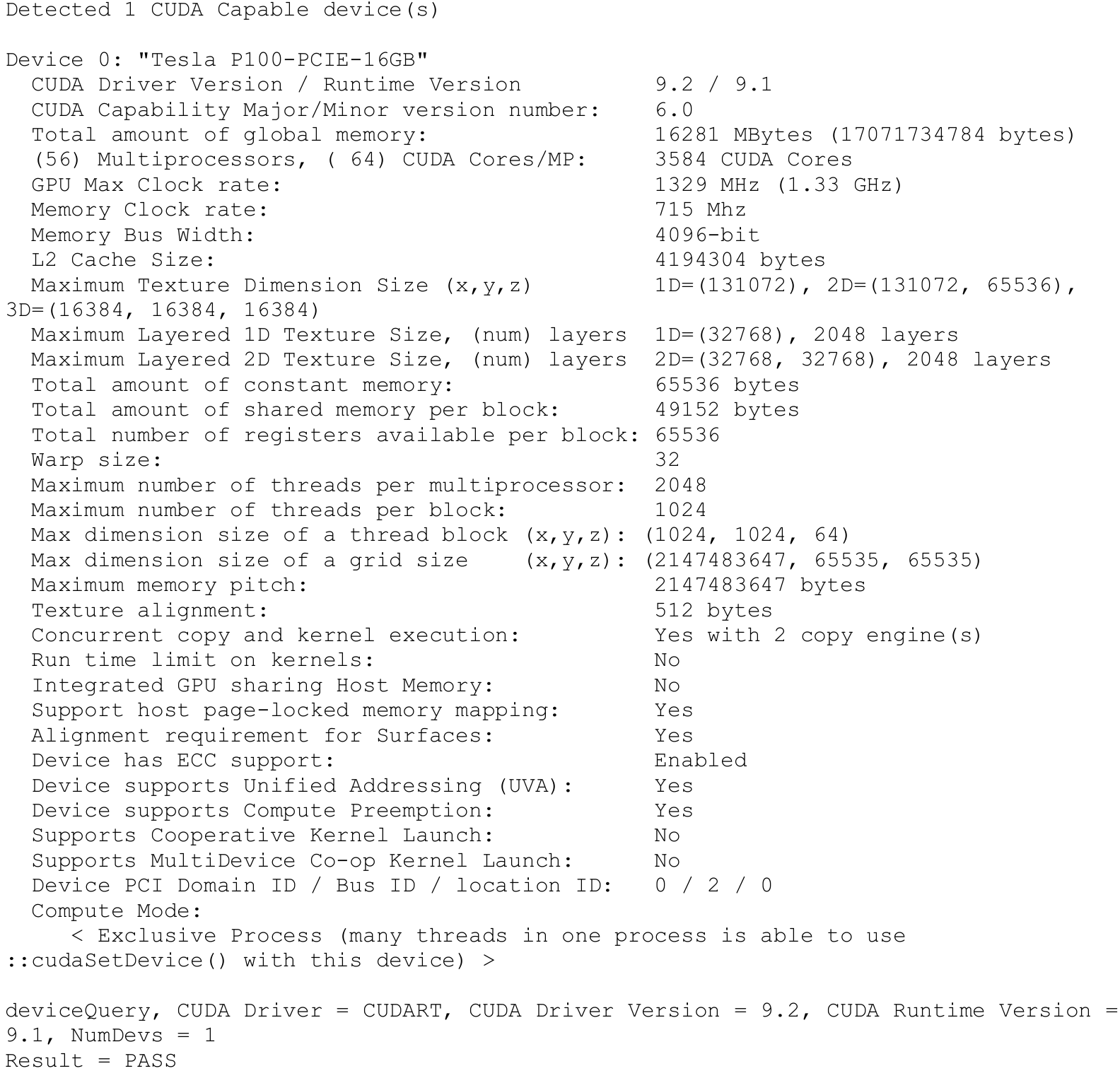}
    \caption{``deviceQuery'' script on an XC50 node}
    \label{fig:deviceQuery}
\end{figure}

Running the CUDA ``deviceQuery'' script on an XC50 node, we get the characteristics of the GPU that it is equipped with (see Figure \ref{fig:deviceQuery}).

\section{Performance Analysis}
\begin{figure}[h]
    \centering
    \includegraphics[scale=0.5]{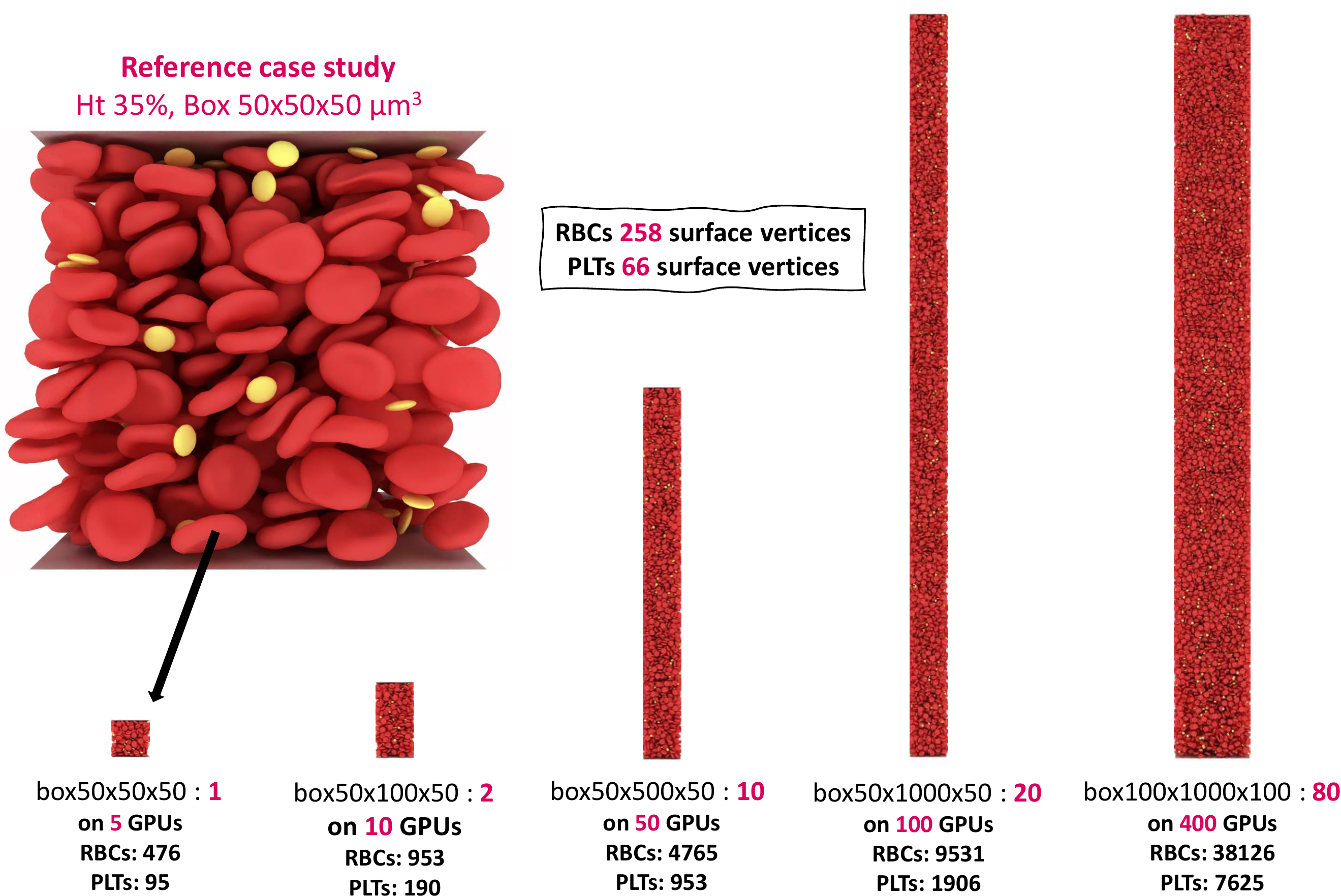}
    \caption{Summary of the weak scaling case studies. For each domain, we present the problem growth in comparison to the reference domain, the number of GPU nodes used to solve the domain and the numbers of blood cells injected in the field. The numbers of blood cells refer to the case of 35\% hematocrit.}
    \label{fig:WeakScalingSummary}
\end{figure}

\begin{figure}[h]
    \centering
    \includegraphics[scale=0.45]{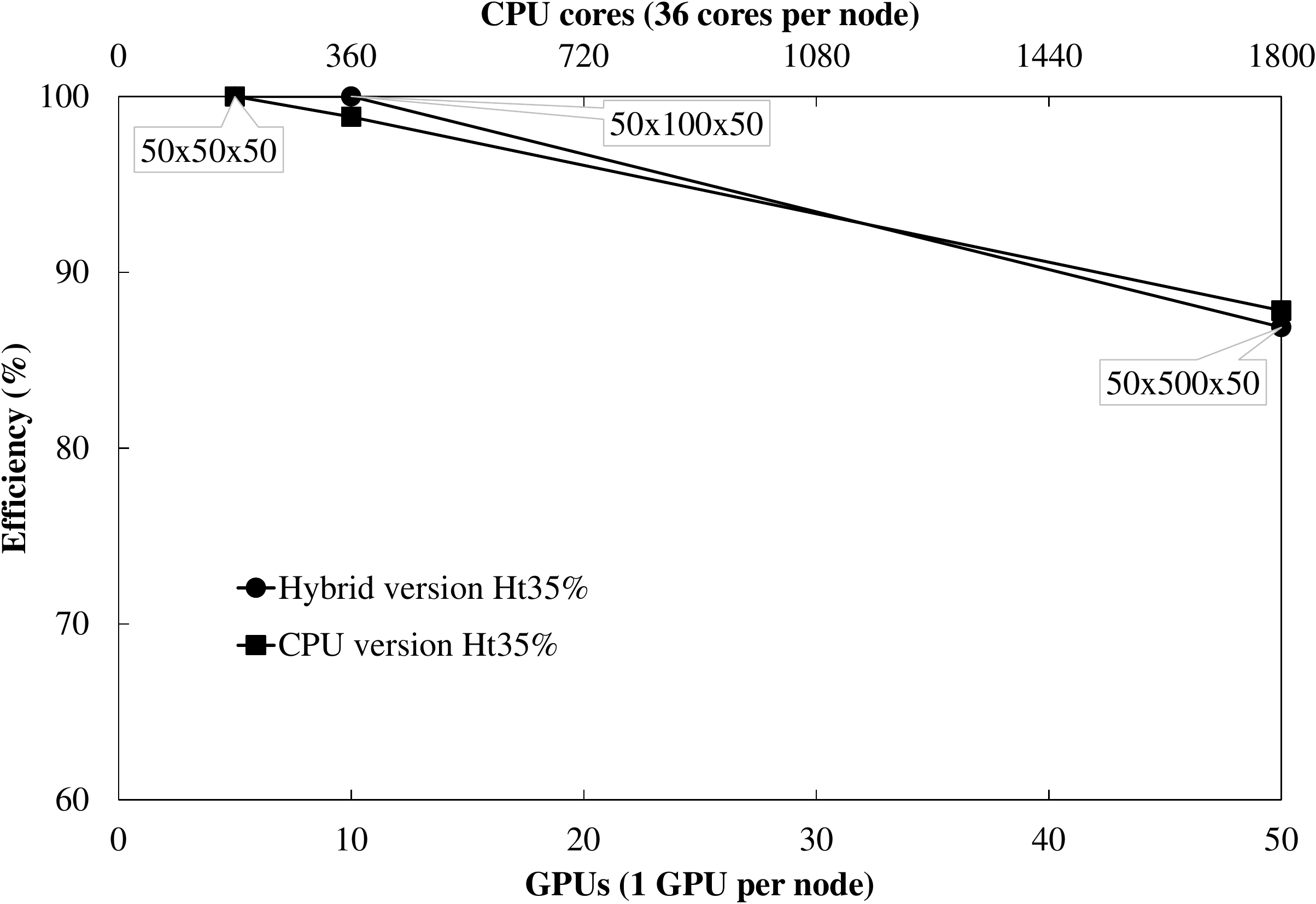}
    \caption{Weak scaling and comparison of the hybrid (CPU/GPU) version with the CPU-only version. The hybrid version is executed on GPU nodes (12 cores and one NVIDIA GPU each), while the CPU-only version on the CPU nodes (36 cores each) of Piz Daint. The upper axis refers to the CPU nodes, while the lower axis refers to the GPU nodes.}
    \label{fig:WeakScalingHybridCPUEfficiency}
\end{figure}

\begin{figure}[h]
    \centering
    \includegraphics[scale=0.45]{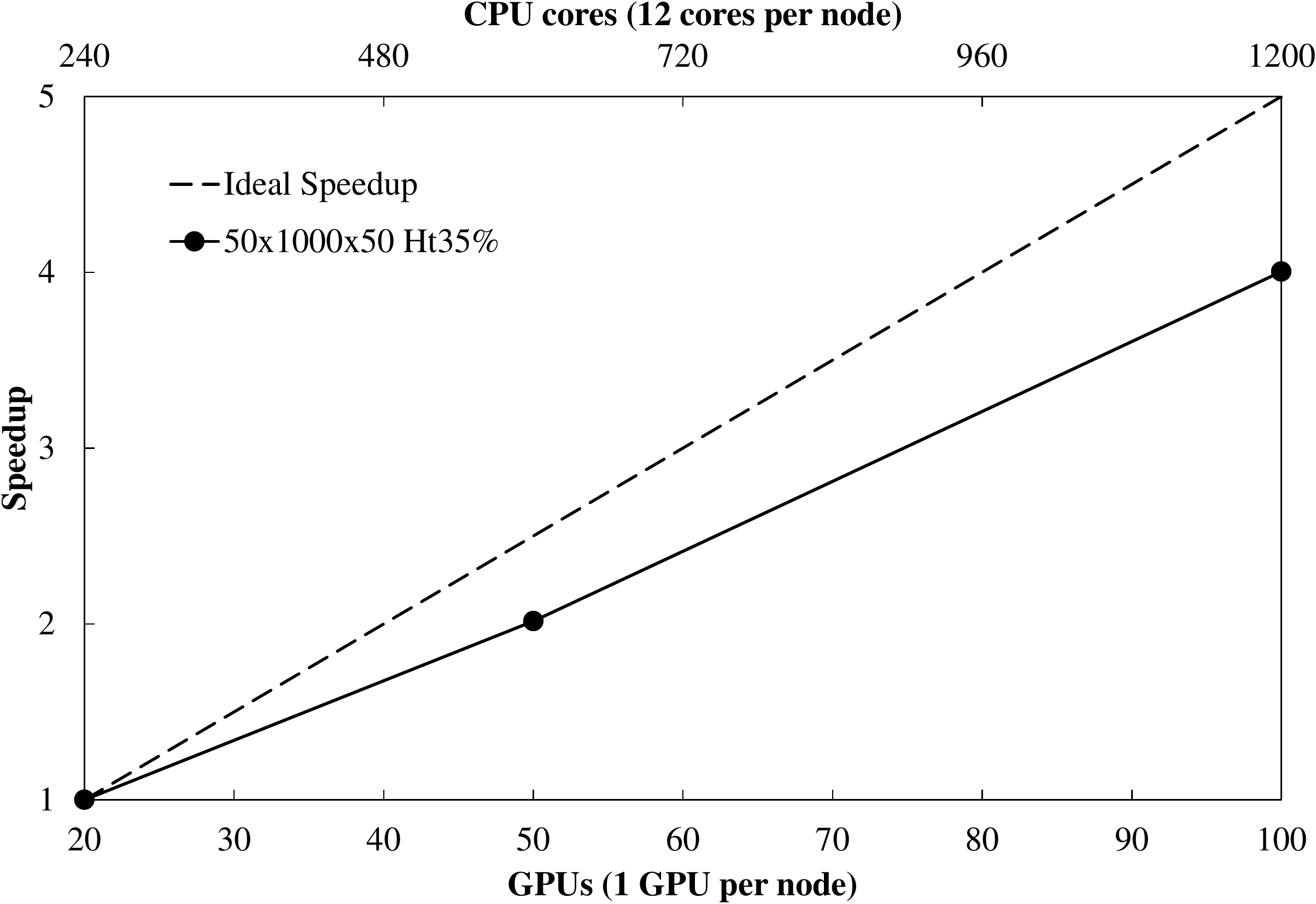}
    \caption{Strong scaling of the hybrid version of the code for a large domain.}
    \label{fig:StrongScalingHybridSpeedup}
\end{figure}

Goal of this section is to complement the performance analysis of the main article. Figure \ref{fig:WeakScalingSummary} summarises the weak scaling case studies (we omit the cases at 45\% hematocrit). Figure \ref{fig:WeakScalingHybridCPUEfficiency} presents the weak scaling comparison between the hybrid version (npFEM on GPUs, LB/IBM on CPUs) and the CPU-only version. Porting part of a library on GPUs can cause extra data management, book-keeping and overhead from new data structures, thus most of the times the efficiency of the ported hybrid library demonstrates a degradation. Obviously (Figure \ref{fig:WeakScalingHybridCPUEfficiency}) this is not the case for our computational framework where we manage to keep the same performance, keeping the overhead at a minimum level. Strong scaling plays a critical role when there is an abundance of computational power. Also, strong scaling partially indicates the scalability of a tool and thus where applicable it is an interesting metric to show. Figure \ref{fig:StrongScalingHybridSpeedup} presents the speedup of the strong scaling for a domain of size 50x1000x50 $\mu m^3$. The speedup is given as $\frac{t_{N_0}}{t_N}$, where $t_{N_0}$ is the time spend in $N_0$ GPU nodes and $t_N$ in $N$ GPU nodes. The speedup is good but not ideal and the reason for this is the extra book-keeping needed for satisfying modularity and genericity of the framework.

\clearpage
\bibliography{References}